\documentclass{jfm}
\usepackage{graphicx}
\usepackage{epstopdf, epsfig}
\usepackage{natbib}
\usepackage[colorlinks=true,citecolor=blue]{hyperref}
\usepackage{float}
\usepackage{tikz}
\usepackage{soul}
\usepackage{makecell}
\usepackage{tikz}
\usepackage{scalerel}
\newcommand\sbullet[1][1.5]{\mathbin{\ThisStyle{\vcenter{\hbox{%
  \scalebox{#1}{$\SavedStyle\bullet$}}}}}%
}

\usepackage{lineno}
\shorttitle{Energy exchanges in rotor wakes}
\shortauthor{N. Biswas, and O.R.H. Buxton}
\title{Energy exchanges between coherent modes in the near wake of a rotor model at different tip speed ratios
}

\author{Neelakash Biswas\aff{1} \corresp{\email{n.biswas20@imperial.ac.uk}},
  Oliver R.H. Buxton \aff{1}
}

\affiliation{\aff{1}Department of Aeronautics, Imperial College London,
London, SW7 2AZ, UK

}

\graphicspath{{figures/}}

\newcommand{\rv}[1]{{\textcolor{black}{#1}}} %  by NB
\newcommand{\rvv}[1]{{\textcolor{black}{#1}}} %  by NB
\newcommand{\rvvv}[1]{{\textcolor{black}{#1}}} %  by NB

\begin{document}

\maketitle

\begin{abstract}
 In this work we investigate the spatio-temporal nature of various coherent modes present in a rotor wake using a combination of new PIV experiments and data from \citet{biswas2024effect}. A multi-scale triple decomposition of the acquired velocity field is sought to extract the coherent modes and thereafter, the energy exchanges to and from them are studied using the multi-scale triple-decomposed coherent kinetic energy budgets developed by \citet{baj2017interscale}. Different frequencies forming the tip vortex system (such as the blade passing frequency, turbine's rotational frequency and their harmonics) are found to be energised by different sources such as production from the mean flow or non-linear triadic interaction or both, similar to the primary, secondary or the mixed modes discussed in \citet{biswas2022energy}. In fact, the tip vortex system forms a complex network of nonlinear triadic energy transfers, the nature and the magnitudes of which depend on $\lambda$. On the other hand, the modes associated with the sheddings from the nacelle or tower and wake meandering are found to be primarily energised by the mean flow. We show that the tip vortex system exchanges energy with the mean flow primarily through the turbine's rotational frequency. In fact, the system transfers energy back to the mean flow through the turbine's rotational frequency at some distance downstream marking the onset location of wake recovery ($x_{wr}$). $x_{wr}$ is shown to reduce with $\lambda$ due to stronger interaction and earlier merging of the tip vortices at a higher $\lambda$.

\end{abstract}


\section{Introduction}
%Coherent structures\\
%What are they\\
%Why important\\

%\citet{nidhan2020spectral} studied the wake of a disk at similar $Re$ using SPOD. They found two modes, in the near wake the vortex shedding mode associated with azimuthal wave number 1 was the most energetic, in the far wake the double helix mode (with azimuthal wave number 2) was more energetic. The vortex shedding mode had a Strouhal number $\approx 0.135$, for the double helix mode it was very low. \citet{berger1990coherent} found an axisymmetric pumping at a low frequency. The vortex shedding mode can persist for very long, $\approx 29D$ \citep{cannon1993observations}. A manifestation of the non-linearity in Naviar Stokes is formation of resonant triads \citep{baj2017interscale, biswas2022energy, schmidt2020bispectral}, POD, problem with POD, DMD, OMD. 

Rotor wakes are inherently multiscale in nature as the flow is simultaneously forced at different time (frequency) and length (wavenumber) scales through the tip vortices, sheddings from the nacelle and the tower and large-scale motions such as wake meandering \citep{abraham2019effect, porte2020wind, biswas2024effect}. All these structures play different roles in the spatio-temporal evolution of the combined rotor wake. There is a consensus that the tip vortices act as a shield in the near field, inhibiting mixing with the background fluid \citep{medici2005experimental, lignarolo2015tip, biswas2024effect}. Active and passive methods have hence been utilised to introduce asymmetry into the helical vortex system to expedite its breakdown process \citep{quaranta2015long, brown2022accelerated, ramos2023multi, abraham2023experimental}, which is a necessary step to initiate the process of wake recovery. 

Although the dynamics of the tip vortices have been the focus of many studies, the importance of the nacelle and the tower in the evolution of the wake has only been realised rather recently \citep{howard2015statistics,pierella2017wind, foti2016wake, de2021pod}. The tower has been shown to act as an important source of asymmetry in the wake by disturbing the tip vortices and hence promoting mixing behind the tower \citep{biswas2024effect, pierella2017wind}. The hub vortex or the shedding behind the nacelle has been linked to the development of wake meandering in the far field, which is associated with large-scale tranverse displacements of the wake centre \citep{howard2015statistics, foti2016wake}. \citet{foti2016wake} showed that the hub vortex formed downstream of the nacelle grows in the radial direction as it moves downstream and interacts with the outer wake thereby potentially augmenting the wake-meandering.  

The dynamics of these length/time scales can be better understood by distinguishing the coherent modes associated with each of them. This can be achieved through a multi-scale triple decomposition of the velocity field \textit{$\textbf{u}(\textbf{x},t)$} (where \textit{\textbf{x}} and $t$ denote space and time respectively in the following form

 \begin{equation}
    \rvv{ \textit{\textbf{u}}(\textit{\textbf{x},t}) = \bar{\textit{\textbf{u}}}  (\textit{\textbf{x}}) + \sum_l 
 \tilde{\textit{\textbf{u}}}_l   (\textit{\textbf{x}},t) + \textit{\textbf{u}}^{''}(\textit{\textbf{x}},t)}
     \label{eq:triple}
 \end{equation}

 Here $\bar{\textit{\textbf{u}}}  (\textit{\textbf{x}})$ is the mean component, $\textit{\textbf{u}}^{''}(\textit{\textbf{x}},t)$ is the stochastic component and $\sum_l 
 \tilde{\textit{\textbf{u}}}_l   (\textit{\textbf{x}},t)$ represents the sum of velocity modes corresponding to individual coherent structures in the flow. This differs from the triple decomposition proposed by \citet{hussain1970mechanics}, where the flow field was decomposed \rvv{into mean, a single periodic (\textit{i.e.} only one characteristic frequency) and fluctuating components,} and hence the simultaneous existence of multiple coherent motions was not addressed. 

 \rv{In previous works, a data-driven approach to extracting the coherent modes in equation \ref{eq:triple} has been taken, typically using a modal decomposition technique \citep{taira2017modal}}. Among them, one of the most commonly used one is Proper Orthogonal Decomposition (POD), where the flow field is decomposed into a series of orthogonal modes that are ranked according to their energy content \citep{lumley1967structure, sirovich1987turbulence}. Despite some limitations, POD has been widely used to identify coherent structures and for reconstruction and modelling of a large variety of flows \citep{taira2017modal}. Another commonly used method is Dynamic Mode Decomposition (DMD), first proposed by \citet{schmid2010dynamic}, which assumes that the time evolution of the flow can be governed by a time invariant, best-fit linear operator \textbf{A} and the eigen-decomposition of the operator gives the so-called DMD modes.  Several variants of the original DMD algorithm have since been proposed with added advantages \citep{schmid2022dynamic}. One amongst them is Optimal Mode Decomposition or OMD proposed by \citet{wynn2013optimal}. The original DMD algorithm obtained the time dynamics by projecting the flow data onto a POD subspace. Contrastingly,  \citet{wynn2013optimal} took a more generalised approach solving a two-way optimisation problem for the flow's dynamics and a low order subspace of \textbf{A}. More details about different modal decomposition techniques can be found in the reviews by \citet{taira2017modal, taira2020modal, schmid2022dynamic}.

 Different modal decomposition techniques such as POD and DMD have been applied to rotor wakes to understand the development and evolution of coherent structures and to develop reduced order models \citep{sarmast2014mutual, debnath2017towards, de2021pod}. \citet{sarmast2014mutual} applied DMD to a LES data set of a wind-turbine wake and found that the dominant modes in the initial phase of tip-vortex evolution agreed well with the predictions of linear stability analysis. \citet{debnath2017towards} performed POD and DMD on a LES data set of a wind-turbine wake with and without a nacelle and tower. For both cases, the dominant mode they reported had a frequency 3 times the rotor's evolution frequency, \textit{i.e.} $3f_r$ (where we denote $f_r$ as the turbine's rotational frequency). Clearly this mode was associated with the tip vortices of the three-bladed turbine. A similar work by \citet{de2021pod} performed POD on a LES data set of the wake of a wind turbine with/without a nacelle and a tower. In the presence of the nacelle and tower, the POD modes in the near field ($x<3.5D$) highlighted the tip vortices (with characteristic frequency $3f_r$), its first super-harmonic (characteristic frequency $6f_r$) and modes associated with vortex shedding from the tower (characteristic frequency $f_T$). \citet{kinjangi2023characterization} applied DMD to a LES data set of a wind-turbine wake that involved a nacelle but no tower. They found modes associated with the turbine's rotational frequency ($f_r$), blade passing frequency ($3f_r$) or the tip vortices, their harmonics and the nacelle's shedding frequency. These results are in line with the recent experiments by \citet{biswas2024effect} that reported a total of six frequencies related to the tip vortices, $f_r - 6f_r$ using Fourier analysis. It was shown that structures with characteristic frequencies such as $f_r$ and $2f_r$ arise in different stages of the merging process of the tip vortices which strongly depends on the tip speed ratio $\lambda$ ($\lambda=\Omega R/U_{\infty}$, where $\Omega$ is the turbine's rotational speed, $R$ is turbine's radius and $U_{\infty}$ is the freestream velocity).

A manifestation of the quadratic non-linearity of the Navier-Stokes equations is the formation of resonant triads \citep{schmidt2020bispectral}. A triad is said to be formed when three frequencies (or wavenumbers) present in the flow sum to zero, \textit{i.e.} $f_1 \pm f_2 \pm f_3=0$. Triadic interactions have been found to play an important role in laminar to turbulence transition \citep{craik1971non, rigas2021nonlinear}, in extreme events such as the formation of rogue waves \citep{drivas2016triad} or intermittent bursts of energy dissipation \citep{farazmand2017variational} or formation of new coherent structures in self-excited turbulent flows \citep{baj2017interscale, biswas2022energy}. The bispectrum (a higher order counter-part of power spectra) has been used to identify such triads in a variety of flows \citep{corke1991mode, schmidt2020bispectral,kinjangi2023characterization, Kinjangi2024}. In fact, \citet{schmidt2020bispectral} introduced a bispectral mode decomposition technique to distinguish modes associated with triadic interactions. \citet{baj2017interscale} showed that such triads exist in the wake of a 2 dimensional array of prisms of different size, generating new frequencies which correspond to the sum/difference of the fundamental shedding frequencies of the different prisms. Using a triple decomposed coherent kinetic energy (CKE) budget equation, they showed that the shedding modes of the prisms were energised primarily by the mean flow henceforth termed them as the `primary modes', while the new frequencies were solely energised by the non-linear triadic interaction term of the CKE budget equation hence identifying them as `secondary modes'. \citet{biswas2022energy} applied similar analysis to a different flow configuration consisting of a cylinder and a control rod and obtained similar energy pathways between the primary and secondary modes. The scenario in a rotor wake is much more complex due to the presence of a large number of modes (as discussed previously) that can form a large number of resonant triads (especially the tip vortices) and due to the three-dimensional nature of the wake. \citet{kinjangi2023characterization} recently performed a similar study on a LES data set of a wind-turbine wake. Although the authors identified the dominant resonant triads in the wake, the direction of energy exchanges to and from the modes was not discussed. The aim of the present work is to extend the works of \citet{baj2017interscale, biswas2022energy} on multiscale cylinder arrays to a rotor wake.

% studied the wake of a wind turbine with POD. Nacelle and tower increases asymmetry in the wake, the turbulence induced by the tower can make the wake recovery faster \citep{de2021pod}, breaks down the tip vortices earlier. \citep{de2021pod} captured the tip vortices and its harmonics using POD. The modes in the near/far wake were different in presence of the tower and nacelle. They found wake meandering mode, but that was most energeyic, probably because the plane they considered was at $y/D=-0.3$ which did not contain the nacelle. 

A large number of time-resolved particle image velocimetry (PIV) experiments are performed on a rotor model incorporating a nacelle and a tower at various tip speed ratios. We use OMD to identify and extract the coherent structures in the flow field. Next, we use the multiscale triple decomposed CKE budget equations derived by \citet{baj2017interscale} to identify the primary energy sources of the various coherent modes. We also quantify the non-linear triadic energy fluxes between different modes forming a triad. Finally we connect the insights gained from the coherent energy budget analysis to the overall wake evolution and wake recovery.

 \begin{figure}
  \centerline{
  \includegraphics[clip = true, trim = 0 0 0 0 ,width= 1\textwidth]{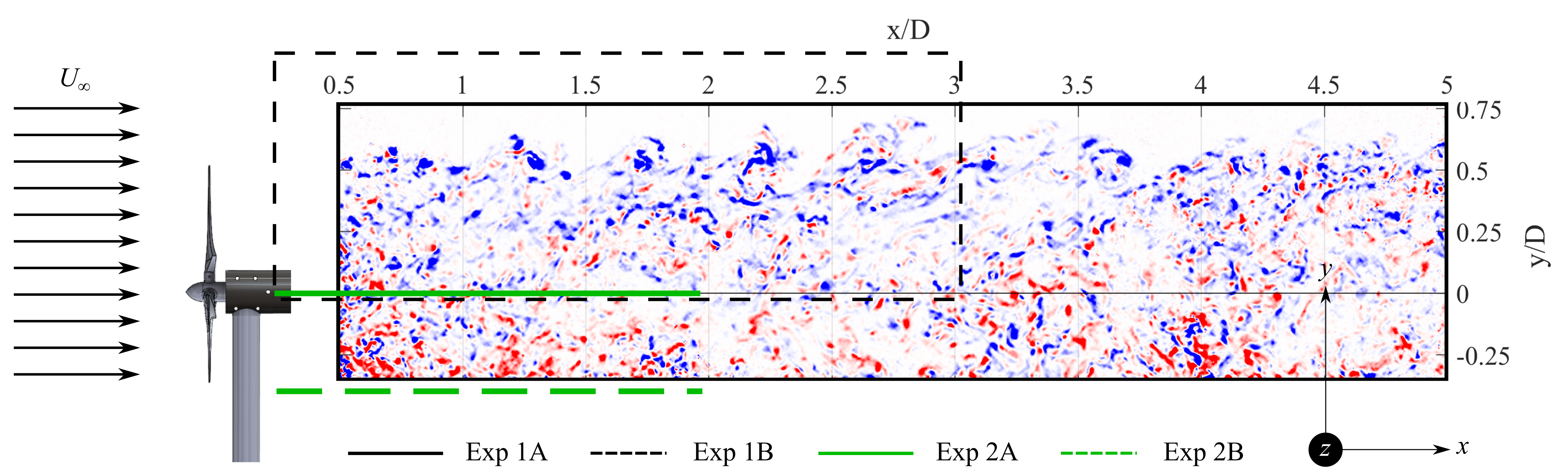}  }
 \caption{Fields of view associated with different PIV experiments. The filled contours show the vorticity field obtained from experiment 1A for $\lambda=6$.}
\label{fig:sch}
\end{figure}

%We consider low inflow turbulence environment, which is a good approximation for off shore wind turbines, where the incoming turbulence levels are generally lower than onshore \citep{ivanell2010stability, lignarolo2015tip}. 

%Meanflow kinetic energy fluxes into the wake increases after the tip vortices undergo leap-frogging \citep{lignarolo2015tip}. 

\section{Experimental method}\label{Section_model}

A large number of PIV experiments were performed on a small-scale rotor in the hydrodynamics flume in the Department of Aeronautics at Imperial College London. The flume had a cross sectional area of $60 \times 60$ cm$^2$ at the operating water depth. The rotor model had a diameter ($D$) of $0.2$ m and was the same as that detailed in \citet{biswas2024effect}. The rotor had a nacelle and tower associated with it such that it resembled a utility-scale turbine. The freestream velocity ($U_{\infty}$) was kept constant at 0.2 m/s and the rotor was driven by a stepper motor to operate it at different tip speed ratios. A total of six tip speed ratios are discussed in this work with the main focus on $\lambda=6$ and $\lambda=5$. Planar PIV experiments were performed in different orthogonal planes. The location and the size of the fields of view associated with different experiments are shown in figure \ref{fig:sch}. Experiment 1 considered the $xy$ plane, where $x$ is the streamwise direction and $y$ is the transverse direction (along the tower's axis). 3 phantom v641 cameras were used simultaneously in experiment 1A giving a field of view (FOV) with a large streamwise extent, up to $x\approx 5D$. Similarly, experiment 1B used 2 cameras, and had a smaller FOV stretching upto $x\approx3D$. Experiment 2 focused on the $xz$ plane, \textit{i.e.} the plane normal to the tower's axis, at different $y$ offsets. For experiment 2A, the laser sheet was aligned with the nacelle's centreline (the solid green line in figure \ref{fig:sch}), while in experiment 2B the sheet was placed $0.35D$ below the nacelle's centreline (the dashed green line in figure \ref{fig:sch}) to capture the tower's wake. Further details about all the experiments are tabulated in table \ref{tab:kd}. For all the experiments, \rvv{images were acquired at an acquisition frequency of $100$ Hz in cinematographic mode (\textit{i.e.} the time between any two successive images was $0.01$s.)} for a total time $T$ $\approx$ 54.5$s$ \rv{($\approx 85$ rotor revolutions for $\lambda=5$)}.

\begin{table}
  \begin{center}
\def~{\hphantom{0}}
  \begin{tabular}{lcccccccc}
      Exp  & $U_{\infty}(m/s)$ & $\lambda$ & FOV & Plane & $\delta x$(mm) & $f_{aq}(Hz)$  & $T(s)$ \\[3pt]
      \vspace{0.15cm}
      
    1A   & 0.2 & 5, 6 &  \hspace{0.2 cm}\makecell{$0.5D<x<5D$,  \\ $-0.35D<y<0.75D$} \hspace{0.2 cm} & $z=0$ & $2.35$ & 100 &  54.75\\
     \vspace{0.15cm}

     1B & 0.2 & \makecell{5.3, 5.5, \\ 6.6, 6.9}  & \hspace{0.2 cm}\makecell{$0.2D<x<3.17D$, \\ $0<y<0.93D$}\hspace{0.2 cm} & {$z=0$}  & $2.07$ & 100 &  54.55\\
\vspace{0.15cm}

   2A & 0.2 &  6 & \hspace{0.2 cm}\makecell{$0.29D<x<1.95D$, \\ $-0.73D<z<0.69D$}\hspace{0.2 cm} & {$y=0$}  & $1.84$ & 100 &  54.55\\
\vspace{0.15cm}
    2B & 0.2 &  6 & \hspace{0.2 cm}\makecell{$0.29D<x<1.95D$, \\ $-0.73D<z<0.69D$}\hspace{0.2 cm} & {$y=-0.35D$}  & $1.84$ & 100 &  54.55\\

  \end{tabular}
  \caption{parameters associated with different experiments}
  \label{tab:kd}
  \end{center}
\end{table}

\section{Coherent modes in the wake}

The vorticity field in figure \ref{fig:sch} shows a large variety of length scales (and associated time scales/frequencies) contained in the rotor wake. \rv{The time evolution of these length scales for different $\lambda$ can be observed in the supplementary videos of \citet{biswas2024effect}. The important scales include} the rotor's rotational frequency ($f_r$), blade passing frequency or the frequency associated with the passage of the tip vortices which is numerically equal to three times the rotational frequency for a three-bladed rotor ($3f_r$). We can also observe the harmonics of $f_r$ and $3f_r$ which although not as energetic as the former, can have an important role in the energy exchange processes as will be discussed. Additionally, there are frequencies associated with vortices shed from the nacelle and the tower which interact with the frequencies related to the tip vortices in a complex fashion \citep{biswas2024effect, de2021pod}. Finally, further downstream, where the near wake transitions to the far wake, the wake meandering frequency can be expected to become important \citep{okulov2014regular,howard2015statistics}. The relative importance of these frequencies and their dependence on $\lambda$ were discussed in detail in \citet{biswas2024effect}.

 \begin{figure}
  \centerline{
  \includegraphics[clip = true, trim = 0 0 0 0 ,width= 1\textwidth]{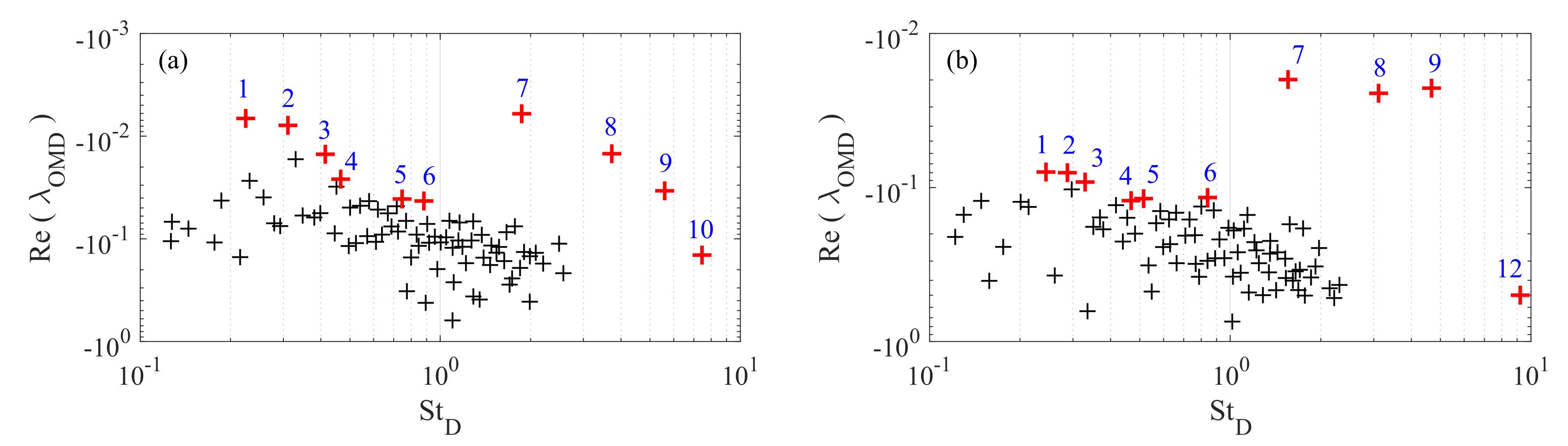}  }
 \caption{OMD spectra obtained for (a) $\lambda = 6$ and (b) $\lambda = 5$ \rvv{from experiment 1A}. \rv{The modes shown by a \textbf{\textcolor{red}{+}} sign are selected for a lower order representation of the flow. The high frequency modes with $St_D>1$ are related to the tip vortices. The low frequency modes ($St_D<1$) are associated with wake meandering, and the sheddings from the nacelle and the tower.}  }
\label{fig:spectra}
\end{figure}

The coherent modes associated with each of these frequencies can be extracted through a multi-scale triple decomposition of the velocity field as described by equation \ref{eq:triple}. The modes corresponding to individual coherent structures in the flow, $\tilde{\textit{\textbf{u}}}_l(\textit{\textbf{x}},t)$ can be obtained using modal decmposition techniques such as Proper Orthogonal Decomposition (POD), Dynamic Mode Decomposition (DMD), Optimal Mode Decomposition (OMD) etc. For the current study we use OMD which is a more generalised version of DMD \citep{wynn2013optimal}. The OMD modes are complex and appear in conjugate pairs (let's denote them as $\phi$ and $\phi^*$). The associated complex time varying coefficients ($a$ and $a^*$) are obtained by projecting the OMD modes back onto the snapshots. Finally, the physical velocity field associated with a mode, \textit{i.e.} $\tilde{\textit{\textbf{u}}}_l(\textit{\textbf{x}},t)$ is obtained through a linear combination of the OMD mode and its coefficient as $\tilde{\textit{\textbf{u}}}_l(\textit{\textbf{x}},t) = a\times \phi + a^* \times \phi^*$. A detailed description of the OMD based multi-scale triple decomposition technique can be found in our previous studies \citep{baj2015triple, baj2017interscale, biswas2022energy}.

 OMD is first performed on the large FOV obtained from experiment 1A. Example OMD spectra are shown in figure \ref{fig:spectra}(a) and \ref{fig:spectra}(b) for $\lambda=6$ and $\lambda=5$ respectively. The rank of the OMD matrices ($r$) was set to 175. A sensitivity study was performed before selecting this $r$ and the results were found to be largely invariant to the selection of $r$. This is discussed in detail in appendix 1. In figure \ref{fig:spectra}, the $x$ axis shows the Strouhal number ($St_D = fD/{U_{\infty}}$) associated with the modes, while the $y$ axis shows the growth rates of the modes. The less damped modes have a growth rate closer to zero and are likely to represent a physically meaningful coherent motion in the flow. These meaningful modes are carefully selected and are highlighted with \textbf{\textcolor{red}{+}} signs in figure \ref{fig:spectra}. Among them the modes on the top right branch of the spectra ($St_D>1$) correspond to the tip vortices and their harmonics. Mode 7 has a frequency equal to the turbines rotation, henceforth denoted as $f_r$. Similarly the other two modes (modes 8 and 9) represent frequencies $2f_r$ and $3f_r$ (blade passing frequency) respectively. \citet{biswas2024effect} showed the presence of higher harmonics of the tip vortices in the flow having frequency up to $6f_r$. However, all the higher harmonics ($4f_r - 6f_r$) could not be captured for a particular $\lambda$ as these are much weaker modes and most of their energy can be expected to be concentrated near the rotor \citep{biswas2024effect}. For $\lambda=6$, only $4f_r$ (mode 10) was captured, while for $\lambda=5$, $6f_r$ (mode 12) could be captured for $r=175$. For $\lambda=6$, OMD was performed with a higher $r=250$ (see appendix 1) but the frequencies $5f_r$ and $6f_r$ were still absent. Any larger rank would overly populate the spectra especially in the low frequency ($St_D<1$) region making it harder to identify the physically meaningful modes. Therefore, $r$ was fixed to 175. To obtain the full set of tip vortex related modes ($f_r - 6f_r$) in the full domain ($x$ up to $5D$), the remaining modes were obtained using phase averaging following \citet{baj2015triple, biswas2022energy}. \rvv{For the low-frequency modes, only six modes were retained with $St_D<1$ that were found to be physically meaningful. The remaining modes in the range $1 \lesssim St_D \lesssim 3$ were found to be much weaker in nature and they did not show any significant energy exchanges. Accordingly, these modes were excluded. A more detailed discussion on this can be found in appendix 2.  }

\begin{figure}
  \centerline{
  \includegraphics[clip = true, trim = 0 0 0 0 ,width= 1\textwidth]{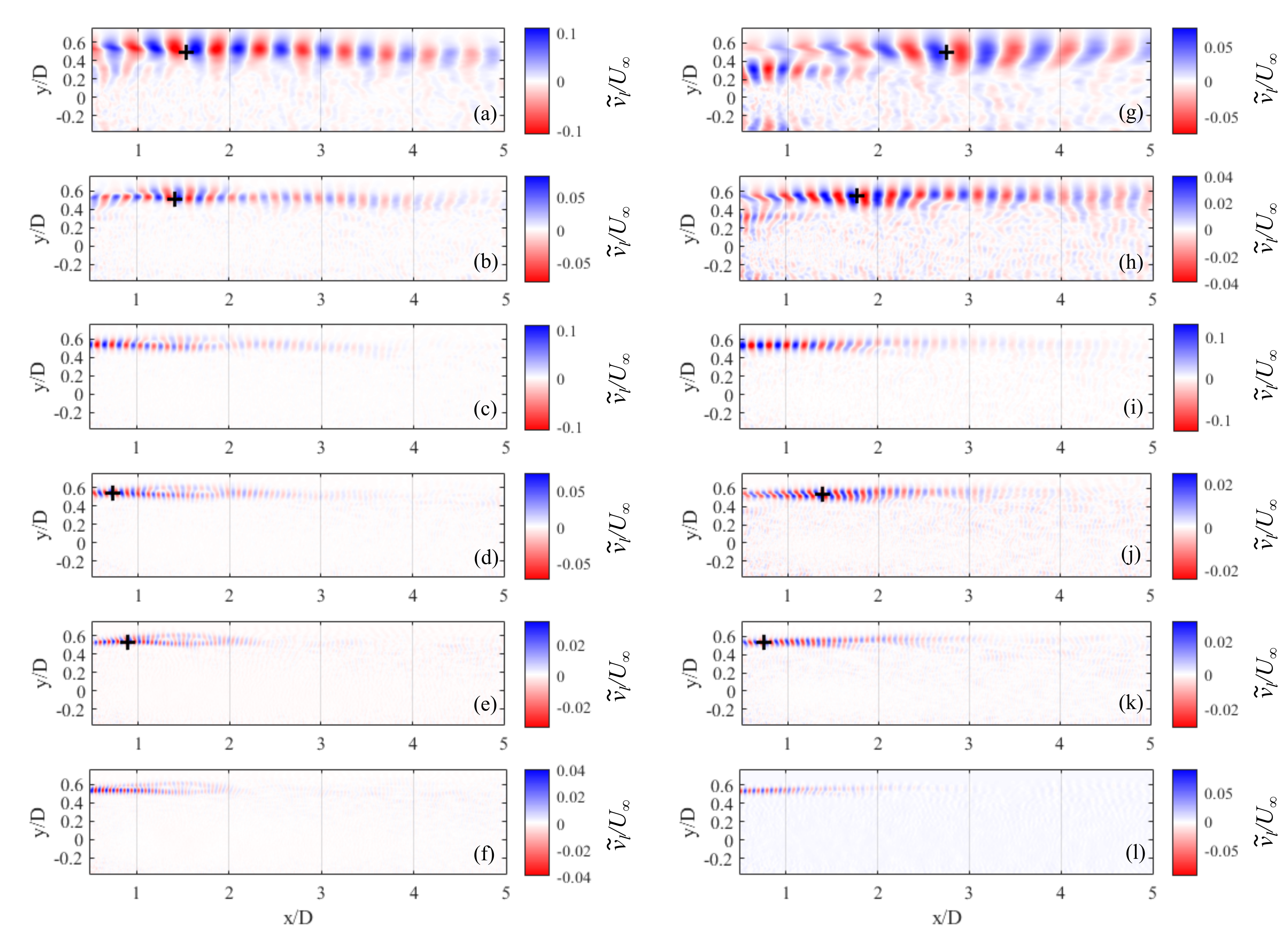}  }
 \caption{Transverse velocity component of the OMD modes associated with $f_r$ - $6f_r$ for $\lambda = 6$ (a - f) and $\lambda = 5$ (g - l). The \textbf{+} sign shows the location where the kinetic energy associated with the individual modes is maximum. }
\label{fig:mode_tip}
\end{figure}

\subsection{Tip vortices}
 
 Let us now look at the spatial nature of the modes associated with the tip vortices obtained from experiment 1A. The transverse velocity components of the modes associated with the frequencies $f_r - 6f_r$ are shown in figure \ref{fig:mode_tip}(a-f) for $\lambda=6$ and in figure \ref{fig:mode_tip}(g-l) for $\lambda=5$. The `\textbf{+}' sign shows the location where the time-averaged kinetic energy of the modes is maximum. Figures \ref{fig:mode_tip}(c) and \ref{fig:mode_tip}(i) show the modes associated with the tip vortices ($3f_r$) and the modes are qualitatively similar for both $\lambda$s. The modes can be expected to be the most energetic near the rotor plane and hence are found to monotonically decay within the field of investigation. The modes associated with the frequency $f_r$ are shown in figures \ref{fig:mode_tip}(a) and \ref{fig:mode_tip}(g) for the two $\lambda$s which represent large-scale structures associated with the merging of the tip vortices \citep{felli2011mechanisms, biswas2024effect}. Note that the spatial organisation of the mode is significantly different for different $\lambda$ unlike the tip vortices. For $\lambda=6$, the mode is stronger and its energy content peaks nearer to the rotor which reaffirms a stronger and earlier interaction between the tip vortices for a higher $\lambda$ \citep{felli2011mechanisms, sherry2013interaction, biswas2024effect}. Furthermore, for $\lambda=5$, there is a region near the root of the blades where $f_r$ is energetic, which is believed to have resulted from an earlier interaction of the unstable root vortices. For $\lambda=6$, the local angle of attack near the root section of the blade does not aid the formation of root vortices. 
 
\rv{ A similar dependence on $\lambda$ is observed for $2f_r$, \textit{i.e.} the mode peaks at an earlier streamwise location for $\lambda=6$ (note the `\textbf{+}' sign), and there is a region near the root where the mode is energetic for $\lambda=5$. Note that $2f_r$ forms a triad with $f_r$ and $3f_r$, suggesting possible triadic energy exchanges between these three modes. Another interesting observation is that the kinetic energy of $2f_r$ peaks at a streamwise location which is between that of $f_r$ and $3f_r$. This is reminiscent of the secondary modes observed in our previous studies for different flow configurations \citep{baj2017interscale, biswas2022energy}. These secondary modes arose from the non-linear triadic interaction between two primary modes of different characteristic frequencies. The downstream streamwise location at which a secondary mode was the most energetic lay between the corresponding locations of the interacting high and low frequency primary modes. Whilst the secondary modes were produced due to triadic interactions, the primary modes were primarily energised by the mean flow. The nature and the origin of the modes we discuss here will be understood in more detail in section \ref{energy} where we will assess the kinetic energy budget associated with each individual mode}, \rvv{but we shall see that similar energy pathways and spatial arrangements exist as in our previous work.} 

The modes associated with $4f_r - 6f_r$ are comparatively weaker and the energy of the modes peaks between the corresponding locations of a number of other modes. For instance, for $\lambda=6$, $5f_r$ peaks between the corresponding locations for $2f_r$ and $3f_r$ and also between that of $f_r$ and $4f_r$, both pairs summing to $5f_r$. Accordingly, a number of modes might contribute to the formation of these high frequency modes. For $\lambda=5$ on the other hand, the peak of $5f_r$ lies only between that of $2f_r$ and $3f_r$, showing an interesting dependence with the tip speed ratio.

 \begin{figure}
  \centerline{
  \includegraphics[clip = true, trim = 0 0 0 0 ,width= 1\textwidth]{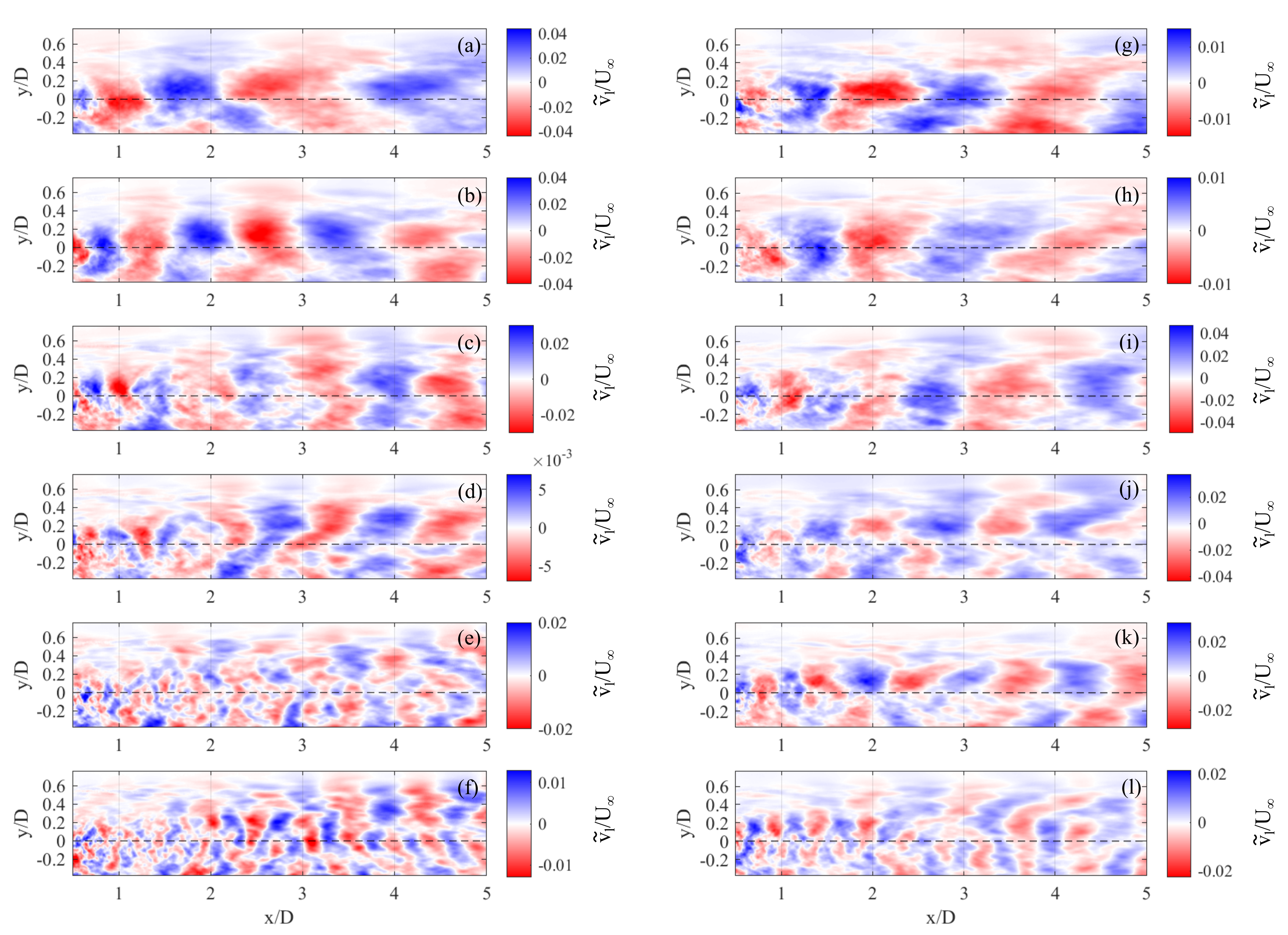}  }
 \caption{Transverse velocity component of the low frequency modes (labeled as 1-6 in figure \ref{fig:spectra}) for $\lambda = 6$ (a - f) and $\lambda = 5$ (g - l). }
\label{fig:mode_inner}
\end{figure}

\subsection{Low frequency modes}
\label{low_f}
Apart from the high-frequency modes related to the tip vortices, there are a number of low-frequency modes observed in the OMD spectra (modes 1-6) in figure \ref{fig:spectra}. For $\lambda = 6$, modes 1 and 2 have $St_D$ around $0.22$ and $0.31$ respectively, which matches well with the range of Strouhal numbers associated with large-scale oscillations due to wake meandering reported in past studies \citep{chamorro2013interaction, okulov2014regular}. Indeed the corresponding transverse velocity fields for modes 1 and 2 presented in figures \ref{fig:mode_inner}(a) and \ref{fig:mode_inner}(b) show large-scale structures similar to wake meandering. For $\lambda = 5$ in figure \ref{fig:spectra}(b), a number of modes are observed in the accepted Strouhal number range for wake meandering. They are shown in figure \ref{fig:mode_inner}(g-i) and they again show large-scale coherence. Note that for both the tip speed ratios, the wake meandering mode starts from close to the nacelle and grows radially in the streamwise direction. The wavelength of the mode is shorter near the nacelle and stretches to around $1.5D-2D$ further downstream which matches well with previous experimental and numerical studies \citep{howard2015statistics, foti2016wake}.

A number of modes are also observed in the OMD spectra at \rvv{$0.4 \lesssim St_D \lesssim 0.5$.}   and \rvv{$0.7 \lesssim St_D \lesssim 0.9$.} . The former when nondimensionalised by the nacelle's diameter instead of turbine diameter yields a Strouhal number around 0.066 -- 0.083 which is similar to the nacelle's vortex shedding frequency reported in previous studies \citep{abraham2019effect, howard2015statistics}. These modes are numbered as modes 3-4 for $\lambda=6$ (figure \ref{fig:spectra}(a)) and as modes 4-5 for $\lambda=5$ (figure \ref{fig:spectra}(b)). These modes are however weaker compared to the wake meandering mode and are not spatially as coherent. A potential reason that the nacelle's shedding was not captured well could be due to the fact that the FOV in experiment 1A did not start \rv{from immediately downstream of the nacelle's rear face}.

\begin{figure}
  \centerline{
  \includegraphics[clip = true, trim = 0 0 0 0 ,width= 1\textwidth]{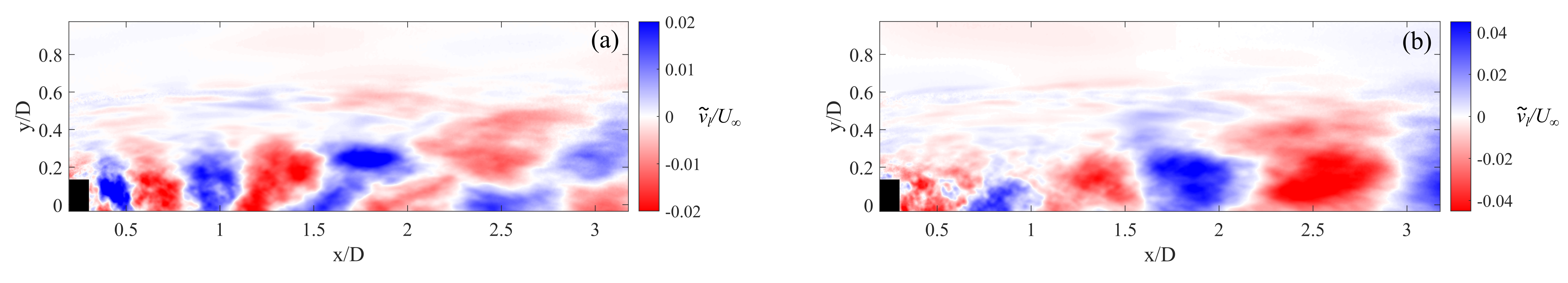}  }
 \caption{Transverse velocity components of the OMD modes associated with frequencies (a) $f_n$ and (b) $f_{wm}$ for $\lambda=5.5$ obtained from experiment 1B. }
\label{fig:mode_fn_fwm}
\end{figure}

Similarly the modes in the Strouhal number range $0.7-0.9$ are most likely related to the shedding from the tower, although the corresponding Strouhal numbers based on the tower's diameter, around $0.074-0.095$ are much lower than the expected value of $\approx$ 0.2 for vortex shedding behind a 2D circular cylinder at a similar Reynolds number $\approx 4000$ based on the tower's diameter \citep{williamson1996vortex}. Such a reduction in the tower's vortex shedding frequency has been observed earlier \citep{de2021pod, biswas2024effect}. \citet{biswas2024effect} argued that a number of factors can play a role such as the reduction of the freestream velocity as the flow passes through the rotor, shear induced on the incoming flow, the unsteadiness in the flow due to the passage of tip and trailing sheet vortices and other three-dimensional effects. As a result the vortex pattern is significantly distorted from the regular vortex street pattern one might expect. The modes which are expected to be associated with the tower's vortex shedding are shown in figures \ref{fig:mode_inner}(e-f) for $\lambda=6$ and in figure \ref{fig:mode_inner}(l) for $\lambda=5$. The modes are much weaker and are not as coherent as the other modes we discussed. This is firstly because of the altered vortex shedding pattern. Secondly, we are only observing the velocity fluctuation parallel to the tower's axis which only arises from the three dimensionality in the vortex shedding pattern and hence is not the dominant velocity component associated with the mode.

Unlike experiment 1A, the FOV of experiment 1B included the rear face of the nacelle (see figure \ref{fig:sch}). OMD was performed for all the $\lambda$s obtained from experiment 1B keeping the rank $r$ fixed to 175. The OMD spectra were similar to those obtained from experiment 1A, consisting of the modes related to the tip vortices and \rvv{an assortment} of low-frequency modes ($St_D<1$). However, the nacelle's shedding mode obtained from experiment 1B was much more coherent than that observed from experiment 1A, as the FOV for the former included part of the nacelle. As an example, the nacelle's shedding mode for $\lambda=5.5$ obtained from experiment 1B is shown in figure \ref{fig:mode_fn_fwm}(a) which shows energetic structures near the nacelle that decay downstream. For a comparison, the wake meandering mode obtained for the same $\lambda$ is shown in figure \ref{fig:mode_fn_fwm}(b) which shows structures of a larger \rvv{spatial extent} that grow downstream and the mode is similar to those observed for experiment 1A.

\begin{figure}
  \centerline{
  \includegraphics[clip = true, trim = 0 0 0 0 ,width= 1\textwidth]{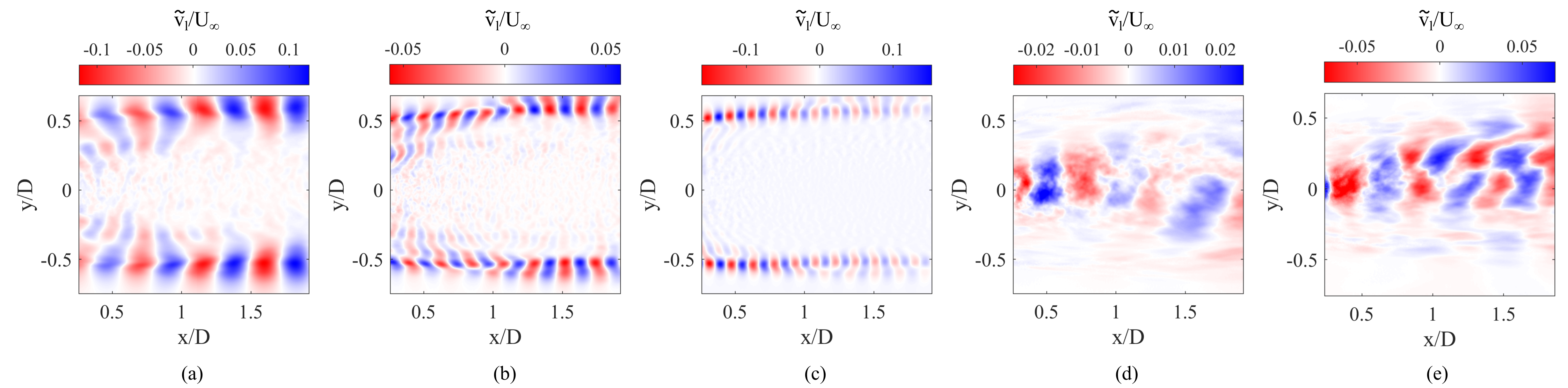}  }
 \caption{Transverse velocity components of the OMD modes associated with frequencies (a) $f_r$, (b) $2f_r$, (c) $3f_r$, (d) $f_n$ obtained in the $xz$ $(y=0)$ plane. Sub figure (e) shows the tower's vortex shedding mode at an offset plane $y=-0.35D$. The modes are shown for $\lambda=6$ only. }
\label{fig:mode_top}
\end{figure}

The OMD modes were also obtained from experiment 2, which considered planes perpendicular to the tower's axis. A selected number of modes are shown in figure \ref{fig:mode_top} for $\lambda=6$. Figures \ref{fig:mode_top}(a-c) show the modes $f_r - 3f_r$ in the $y=0$ plane which are similar to those observed in figures \ref{fig:mode_tip}(a-c). In the same plane, figure \ref{fig:mode_top}(d) shows the nacelle's shedding mode with a Strouhal number based on nacelle's diameter of around 0.069. The tower's shedding mode was not captured in the $y=0$ plane. However, it was captured in the offset plane ($y=-0.35D$) and is shown in figure \ref{fig:mode_top}(e). The mode had a Strouhal number based on tower's diameter of around $0.08$. Note that the mode is much more organised spatially and is more energetic than that observed in the $xy$ plane. Interestingly, no modes could be captured from experiment 2 that resembled the wake meandering modes observed in figure \ref{fig:mode_inner}. \rvv{This is probably because the streamwise extent (up to $x/D \approx 1.85D$) of the field of view was not large enough to capture the structures associated with wake meandering which can have wavelengths as large as $1.5D-2D$ \citep{howard2015statistics}. Additionally, the wake meandering mode can be expected to be more energetic beyond $x/D \approx 2$ (see figure \ref{fig:mode_fn_fwm}(b)), making it highly unlikely to be captured in the short field of view}.

\section{Energy exchanges}
\label{energy}
The energy exchanges to and from the coherent modes can be explored using the multiscale triple decomposed coherent kinetic energy (CKE) budget equations developed by \citet{baj2017interscale}. The CKE budget ($\tilde{k}_l$) equation can be represented in symbolic form as

\begin{equation}
    \textcolor{black}{\frac{\partial \tilde{k}_l}{\partial t}   = -\textcolor{black}{\tilde{C}_l} + \tilde{P}_l - \textcolor{black}{\hat{P_l}} +\textcolor{black}{\Big( \tilde{T}^+_l - \tilde{T}^-_l\Big)} -  \textcolor{black}{\tilde{\epsilon_l}} +\textcolor{black}{\tilde{D_l} }}
    \label{eqn:coherent_TKE}
\end{equation}

In equation \ref{eqn:coherent_TKE}, the source terms on the right hand side consist of convection ($\bar{C}_l$), production from the mean flow ($\tilde{P}_l$), production of stochastic turbulent kinetic energy directly from coherent mode $l$ ($\hat{P}_l$), triadic energy production ($\tilde{T}^+_l - \tilde{T}^-_l$), direct dissipation from coherent mode $l$ (${\tilde{\epsilon}}_l$) and diffusion ($\tilde{D}_l$). The full composition of each of these terms is available in \citet{baj2017interscale}. \citet{baj2017interscale} showed that the triadic energy production term can become significant only when there exists three frequencies that linearly combine to zero or in other words, there is a triad in the form

\begin{equation}
    f_l \pm f_m \pm f_n = 0
\end{equation}

\noindent \citet{baj2017interscale, biswas2024effect} reported the existence of such triads in two different flow configurations involving two dimensional cylinders of unequal diameters. Note that for the present case involving a rotor wake, the frequencies $f_r - 6f_r$ form a large number triads, implying the triadic energy exchange term can play a significant role. We first assess the kinetic energy budgets of the modes obtained from experiment 1A for the tip speed ratios 6 and 5. As we only have planar data, we have to ignore the terms containing out of plane velocity and velocity gradients in the energy budget equation. \rv{We also have to ignore the contribution to the diffusion term from the pressure.} However, as we will show this simplification does not alter the conclusions of the energy budget analysis. For both $\lambda=6$ and $\lambda=5$, the 12 modes shown in figures \ref{fig:mode_tip} and \ref{fig:mode_inner} are selected and the stochastic component of the flow is obtained by subtracting these modes from the mean-subtracted velocity field. \rvv{It is worthwhile mentioning that this process leaves the spectrum of the stochastic turbulence continuous, and reminiscent of classical homogeneous turbulence \citep{baj2015triple, baj2017interscale}}.

 \begin{figure}
  \centerline{
  \includegraphics[clip = true, trim = 0 0 0 0 ,width= 1\textwidth]{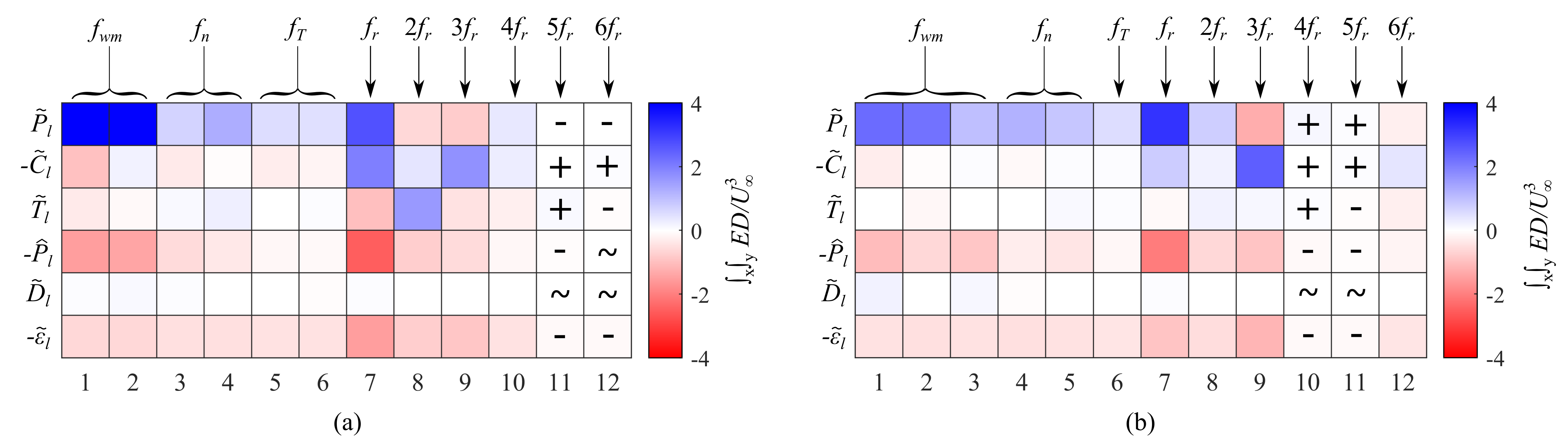}  }
 \caption{\rvv{Energy budget terms of equation \ref{eqn:coherent_TKE}} summed over the domain of investigation for different modes for (a) $\lambda=6$ and (b) $\lambda=5$. }
\label{fig:heat_all}
\end{figure}

For an overall understanding, we first integrate the terms from the CKE budget over the entire domain of investigation for all the modes. The results are shown in figures \ref{fig:heat_all}(a) and \ref{fig:heat_all}(b) for $\lambda=6$ and $\lambda=5$. For the low-frequency modes ($St_D<1$) including the wake meandering and nacelle or tower's vortex shedding, the primary energy source is the $\tilde{P_l}$ term or energy production from the mean flow, similar to the primary modes discussed in \citet{baj2017interscale, biswas2022energy}. Among these primary modes, the wake meandering mode draws the highest amount of energy from the mean flow for both $\lambda$s. The wake meandering mode for $\lambda = 6$ is more strongly energised than for $\lambda=5$. A similar dependence of the strength of the wake meandering mode on $\lambda$ was reported in \citet{biswas2024effect}. \rvv{They established a link between $\lambda$ and the effective porosity of the turbine. As $\lambda$ increased, the effective porosity reduced, resulting in a decrement in the wake meandering frequency and an increment in the strength of the mode.} Note from figure \ref{fig:spectra} that for the low-frequency modes, multiple triads are possible for which the frequencies sum to $\approx 0$. For instance, for $\lambda=6$, $f_1+f_1-f_4 \approx 0$, where $f_1$ is the mode numbered as 1 or the wake meandering mode ($f_{wm}$) and similarly $f_4$ is the 4th mode in the spectrum which is believed to be related to the nacelle's shedding ($f_n$). Another possible triad is $f_3+f_4 - f_6 \approx 0$, where $f_3$ and $f_4$ are related to $f_n$ and $f_6$ is likely related to the shedding from the tower ($f_T$). However, the triadic energy production term was not found to be the dominant term for any of these low-frequency modes as can be seen from figure \ref{fig:heat_all}. The modes associated with $f_n$ are found to be slightly energised by the $\tilde{T}^+_l - \tilde{T}^-_l$ term, while the wake meandering mode loses some energy due to non-linear interactions. All the low-frequency modes lose energy primarily through dissipation and production of incoherent (stochastic) turbulence. Hence, we can say that non-linear triadic interactions are possible among these modes, however they are not the driving feature of their dynamics. 

The tip vortices on the other hand are energised by a variety of energy sources. For both $\lambda$s,  $f_r$ is primarily energised by the mean flow, thus behaving similarly to a primary mode. $3f_r$ only shows a positive convection (-$\tilde{C}_l$) term. \rv{This is expected as the tip vortices are formed at the passage of the blades and then advected into the PIV domain where they decay monotonically}. We can expect a contribution from pressure diffusion to energise the mode, however, this cannot be captured with the current experiments. The nature of $2f_r$ changes with $\lambda$. For $\lambda=6$, it is solely energised by the non-linear triadic interaction term, hence acting like a secondary mode. \rvv{For $\lambda=5$, it is energised primarily by the mean flow production term, while there is also some contribution from the nonlinear triadic interaction term. The mode therefore acts like a `mixed mode', first reported by \citet{biswas2024effect}}. The modes $4f_r - 6f_r$ are weaker compared to $f_r - 3f_r$. \rv{Specifically, $5f_r$ and $6f_r$ for $\lambda=6$ and $4f_r$ and $5f_r$ for $\lambda=5$ exhibited only very weak energy exchanges making it hard to see them in figure \ref{fig:heat_all}}. Therefore a `\textbf{+}' or `\textbf{-}' sign is added to represent gain or loss of energy for these modes. A `\textbf{$\sim$}' sign is shown when the contribution from  a term is found to be negligible ($|\int_x \int_y ED/U_{\infty}^3| <0.01$). Note that for both $\lambda$s $6f_r$ behaves similarly to $3f_r$. For $\lambda=6$, $4f_r$ and $5f_r$ behave similarly to $f_r$ and $2f_r$ respectively. For $\lambda=5$ on the other hand, $4f_r$ behaves like $2f_r$ and $5f_r$ behaves like $f_r$.

\begin{figure}
  \centerline{
  \includegraphics[clip = true, trim = 0 0 0 0 ,width= 1\textwidth]{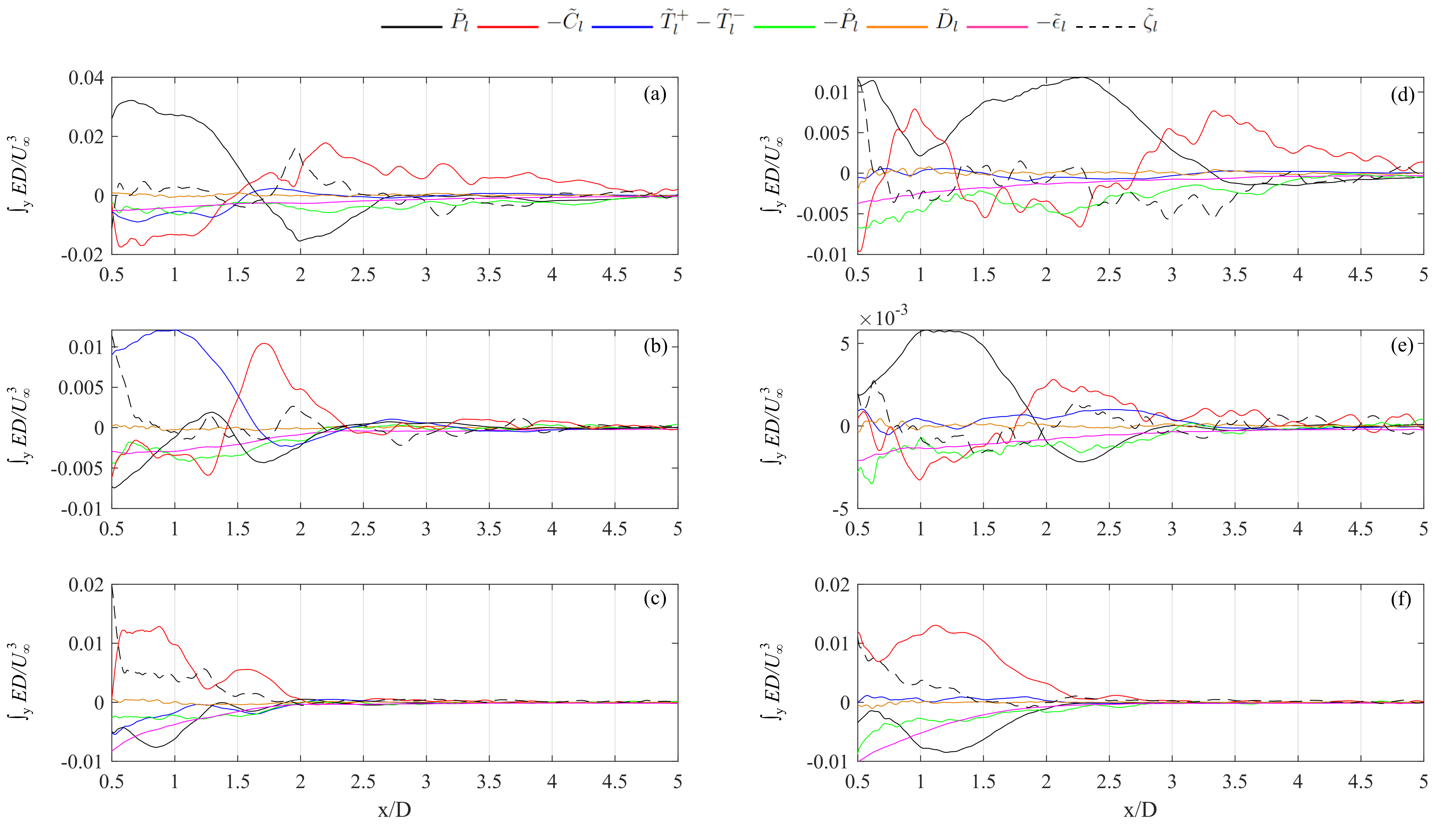}  }
 \caption{Streamwise evolution of the spanwise (along $y$) averaged \rvv{energy budget terms of equation \ref{eqn:coherent_TKE}} for (a) $f_r$, (b) $2f_r$ and (c) $3f_r$ for $\lambda=6$. Sub figures (d)-(f) show the same for $\lambda=5$. }
\label{fig:tip_budget_line}
\end{figure}

For a deeper understanding of the energy budget terms we take a spanwise (along the $y$ direction) \rvv{integral} of the budget terms and look at its streamwise variation. This is first shown for the modes $f_r - 3f_r$ for the two $\lambda$s in figure \ref{fig:tip_budget_line}. \rvv{The corresponding plots for $4f_r - 6f_r$ are not shown as their budgets were similar to one of the first three modes ($f_r - 3f_r$) as discussed earlier}. Note that we consider the budgets in a time-averaged sense so $dk/dt$ is essentially zero, \rv{therefore, the combined effect of the various terms of the CKE budget equation is reflected in the convection term ($-\tilde{C_l}$): if the energy content of the mode is increasing with downstream distance (it is being net energised) then this term will be negative whilst it will be positive when the mode is spatially decaying. }

For $f_r$ (figure \ref{fig:tip_budget_line}(a) and \ref{fig:tip_budget_line}(d)), the primary source is the $\tilde{P}_l$ term, shown with a black line. Note that for $\lambda=6$, the $\tilde{P}_l$ term changes sign at $x/D \approx 1.65$. This location is close to where $f_r$ was found to be the most energetic (see figure \ref{fig:mode_tip}(a)). Beyond this point the mode decays monotonically as indicated by the convection ($-\tilde{C_l}$) term being positive. For brevity, let's denote the location where $\tilde{P_l}$ changes sign as $x_{wr}$. Note that $x_{wr}$ is particularly important in terms of wake recovery, as beyond this point the mode transfers energy back to the mean flow. For $\lambda=5$ (figure \ref{fig:tip_budget_line}(d)), $x_{wr} \approx 3.2$ which implies that wake recovery starts much later. Note that for $\lambda=5$, there is a drop in the $\tilde{P}_l$ term of $f_r$ at $x/D \approx 1$. This is because of the presence of the merged root vortices (with frequency $f_r$) that decay in this region and \rv{noting that the $\tilde{P}_l$ term is essentially the sum of contributions from both the root and tip regions. }

\begin{figure}
  \centerline{
  \includegraphics[clip = true, trim = 0 0 0 0 ,width= 0.9\textwidth]{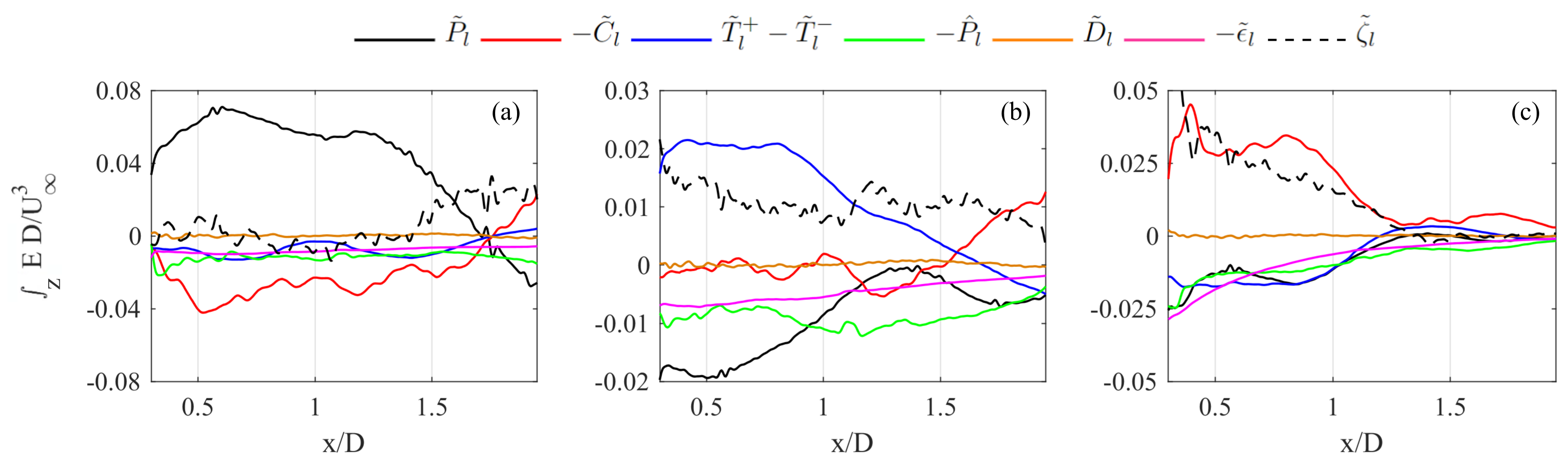}  }
 \caption{Streamwise evolution of the \rvv{energy budget terms of equation \ref{eqn:coherent_TKE}} averaged along the $z$ direction for (a) $f_r$, (b) $2f_r$ and (c) $3f_r$ for $\lambda=6$.}
\label{fig:tip_budget_line_top}
\end{figure}

\begin{figure}
  \centerline{
  \includegraphics[clip = true, trim = 0 0 0 0 ,width= \textwidth]{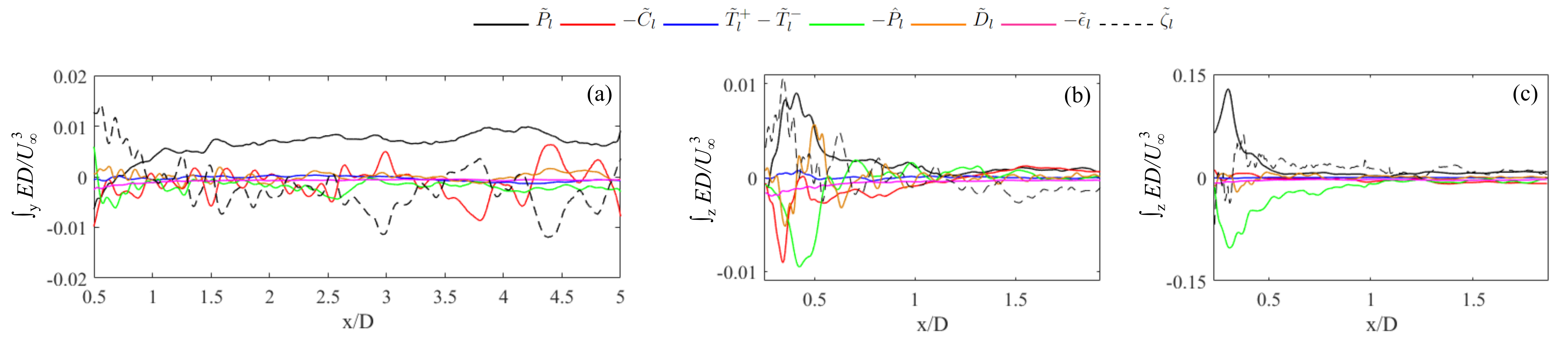}  }
 \caption{Streamwise evolution of the spanwise (along $y$) averaged \rvv{energy budget terms of equation \ref{eqn:coherent_TKE}} for (a) $f_{wm}$ in the $xy$ plane. Streamwise evolution of the energy budget terms averaged along $z$ direction for (b) $f_n$ in the $y=0$ plane, (c) $f_T$ in the $y=-0.35D$ plane.}
\label{fig:inner_budget_line}
\end{figure}

For $\lambda=6$, $2f_r$ is primarily energised by the triadic interaction term which drops off to zero at $x/D \approx 1.6$; this is again close to the point where $2f_r$ is most energetic in figure \ref{fig:mode_tip}(b). From figure \ref{fig:mode_tip} we can see that for $\lambda=5$, $2f_r$ is much weaker than for $\lambda=6$. This is corroborated by the fact that the magnitude of the energy budget terms of $2f_r$ is smaller for $\lambda=5$, compared to $\lambda=6$. Additionally, for the lower $\lambda$, the triadic interaction term is much weaker due to increased separation between the tip vortices. Therefore, unlike $\lambda=6$, for $\lambda=5$, $2f_r$ is thus energised mainly by the mean flow. For $3f_r$, the convection term is positive throughout as it decays monotonically in the domain. The trends are qualitatively similar for both the $\lambda$s. Note also that the residuals of the budget equation are highest for $3f_r$, especially closer to the rotor. This is probably because there is a significant role of the pressure diffusion term in this region that has been neglected in the analysis.   

The same analysis is performed with the data from experiment 2A (see table \ref{tab:kd} for details) which considered the orthogonal $xz$ plane for $\lambda=6$. The budget terms are summed in the $z$ direction and are shown in figure \ref{fig:tip_budget_line_top} for the modes $f_r - 3f_r$. Note that the trends of the budget terms are quite similar to those observed in the $xy$ plane earlier. The dominant source terms of the modes are the same. Moreover, the $\tilde{P}_l$ term for $f_r$ changes sign at the same location as before, at $x/D \approx 1.65$. Similarly, the triadic interaction term for $2f_r$ changes sign at $x/D \approx 1.6$, consistent with the observation in the $xy$ plane. This similarity of the budget terms in the two orthogonal planes shows that the budget terms are close to axisymmetric in nature and \rv{offers reassurance as to the repeatability of our results.}

\begin{figure}
  \centerline{
  \includegraphics[clip = true, trim = 0 0 0 0 ,width= 0.8\textwidth]{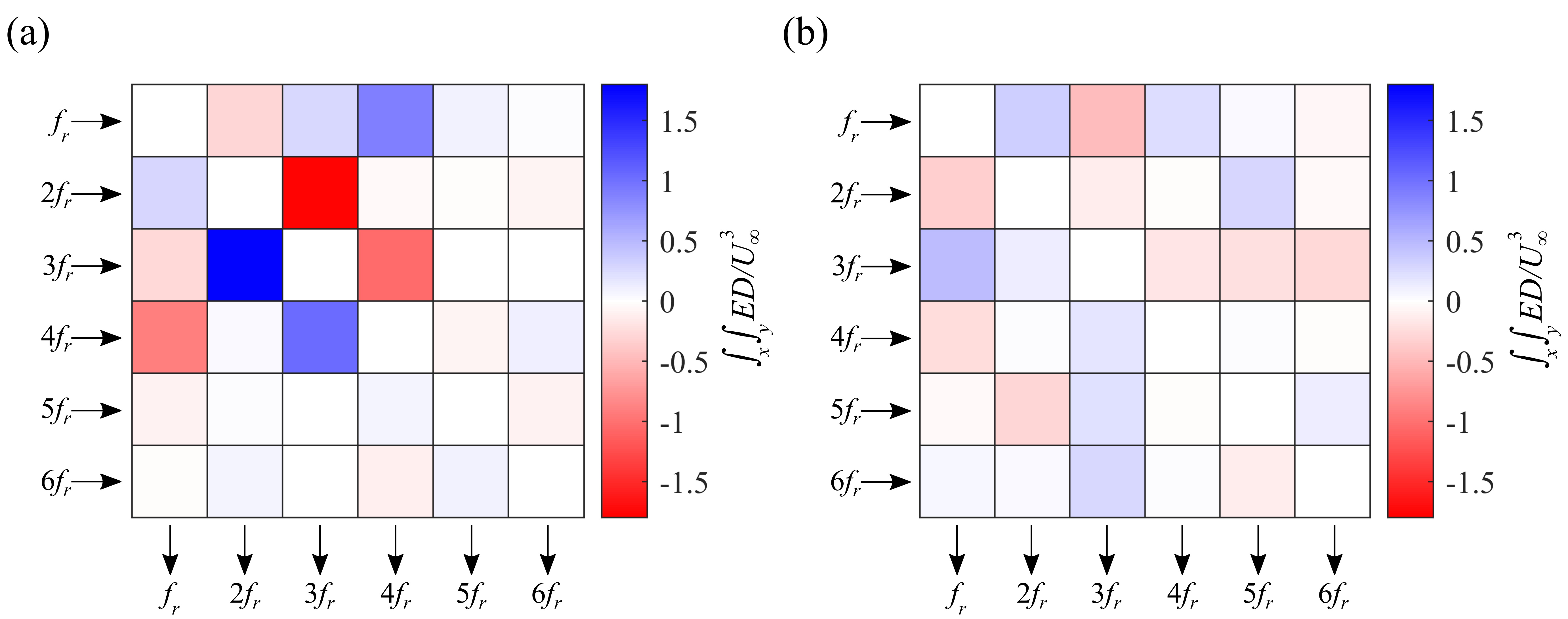}  
 
  }
 \caption{ \rvvv{Triadic energy exchanges for (a) $\lambda=6$ and (b) $\lambda=5$ from experiment 1A.} }
\label{fig:triad_full}
\end{figure}

\rvv{Figure \ref{fig:inner_budget_line} shows the evolution of the CKE budget terms for selected low-frequency modes in different PIV planes for $\lambda=6$. Figure \ref{fig:inner_budget_line}(a) shows the terms for the dominant wake meandering mode obtained from experiment 1A (the mode shown in figure \ref{fig:mode_inner}(a)). Figures \ref{fig:inner_budget_line}(b) and \ref{fig:inner_budget_line}(c) show the same for the nacelle's and the tower's shedding mode obtained from experiments 2A and 2B respectively. The terms appear to be noisier compared to the tip vortex system, however, the $\tilde{P}_l$ term is unambiguously the dominant source term for these modes. For the wake meandering mode, $\tilde{P}_l$ slowly grows with streamwise distance, which is consistent with the observed far-wake dominance of wake meandering \citep{biswas2024effect}. The sheddings from the nacelle and the tower on the other hand quickly drop to zero, showing their relatively transient spatial nature. }

\subsection{Triadic interactions}

Let us now cast a closer look at the triadic energy exchange term. So far we only know that for $\lambda=6$, $2f_r$ is energised primarily by the triadic energy exchange term. But, $2f_r$ forms a number of triads and the triadic energy gain of $2f_r$ is a sum of contributions from all the triads that $2f_r$ can form with the other frequencies. At this point we can ask are there any particular triads that are more important or are there any frequencies that transfer more energy to $2f_r$? Answering these questions can help significantly simplify what would be a rather complicated network of energy transfers. The terms $\tilde{T}^+_l$ and $\tilde{T}^-_l$ in equation \ref{eqn:coherent_TKE} indicate the net non-linear energy gain and loss respectively from the $l_{th}$ coherent mode and are defined as follows:

\begin{equation}
    \tilde{T}^+_{l} = -\frac{1}{2}\sum_{f_s,f_t}\overline{\tilde{u}^{f_l}_i \tilde{u}^{f_t}_j \frac{\partial \tilde{u}^{f_s}_i}{\partial x_j}}, \hspace{1cm} \tilde{T}^-_{l} = -\frac{1}{2}\sum_{f_s,f_t}\overline{\tilde{u}^{f_s}_i \tilde{u}^{f_t}_j \frac{\partial \tilde{u}^{f_l}_i}{\partial x_j}}.
\end{equation}

Note that the terms consist of 3 frequencies and are non-negligible only when the frequencies form a triad \citep{baj2017interscale}. Essentially we take contributions from all possible triads by summing over the two frequencies ($f_s$ and $f_t$). \rvvv{Instead, we can fix one of these two frequencies (let's say $f_s$) and can take a summation over the other frequency ($f_t$).} This would give us the net triadic energy exchange between $f_l$ and $f_s$. We do this for all the frequencies related to the tip vortices ($f_r - 6f_r$) and show the results in figure \ref{fig:triad_full} for the two $\lambda$s. The arrows show the direction of energy transfer. For instance the first row shows the net amount of energy $f_r$ is giving away to the other frequencies. In other words, the first column shows the amount of energy received by $f_r$ from all other frequencies. As observed earlier, the magnitude of the traidic energy transfers are much stronger for $\lambda=6$, compared to $\lambda=5$. \rvvv{Moreover, for $\lambda=6$, most of the energy transfer is limited between $f_r - 4f_r$, while the contributions from $5f_r$ and $6f_r$ are significantly weaker in comparison}. Looking at the second column in figure \ref{fig:triad_full}(a) we can say that the main source for $2f_r$'s triadic energy gain is $3f_r$. \rv{Thereafter, $2f_r$ transfers some amount of energy to $f_r$, hence forming a net inverse energy cascade}. Note that this sequence of energy transfer is similar to that predicted by \citet{felli2011mechanisms} (see figure 41(b) of \citet{felli2011mechanisms}) for a three bladed turbine. \rv{However, the fact that the higher harmonics can also have an important role in the energy exchange process was not highlighted in their work. Note from figure \ref{fig:triad_full}(a) that $f_r$ transfers a significant amount of energy to $4f_r$ and $4f_r$ transfers almost the same amount of energy to $3f_r$. Therefore, although $4f_r$ does not gain energy through triadic interactions, it forms a bypass route of energy transfer allowing $f_r$ to transfer some energy back to $3f_r$}. 

%The same analysis is performed with data from experiments focusing on different planes and experiments with different spatial resolutions for $\lambda=6$. The same pattern of inter-frequency energy transfer is found for $\lambda = 6$ for all the experiments apart from a slight variation in the magnitude of the transfers which confirmed that this particular pattern of energy transfer is a robust feature of the tip vortices at this particular $\lambda$. This is discussed in more detail in appendix 1. 

\begin{figure}
  \centerline{
  \includegraphics[clip = true, trim = 0 0 0 0 ,width= \textwidth]{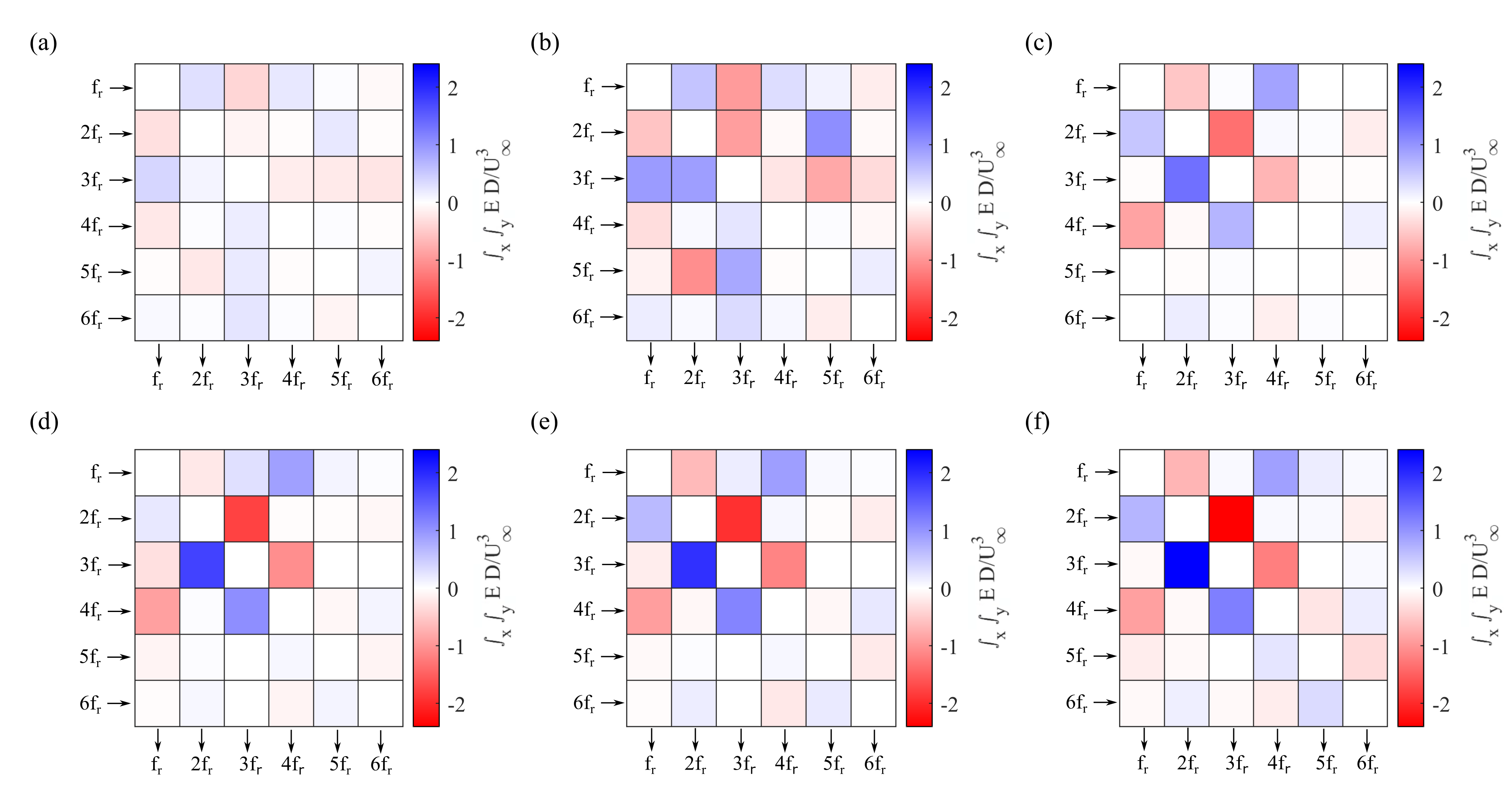}  
 
  }
 \caption{ Net traidic transfers for (a) $\lambda = 5$, (b) $\lambda = 5.3$, (c) $\lambda = 5.5$, (d) $\lambda = 6$, (e) $\lambda = 6.6$, (f) $\lambda = 6.9.$ }
\label{fig:triad_lam}
\end{figure}

For $\lambda=5$, figure \ref{fig:triad_full}(b) shows that not only are the magnitudes of the energy transfers weaker but the energy transfer pathways are also different. This includes the fact that the net non-linear energy gain for $2f_r$ is now quite small, as was previously shown in figure \ref{fig:heat_all}(b). Unlike for $\lambda=6$, $2f_r$ and $4f_r$ are both energised mainly by $f_r$, \rv{as far as triadic interaction is concerned}. To further understand the dependence of these energy transfers on $\lambda$, the analysis is repeated for other $\lambda$s in experiment 1B. These are shown in figure \ref{fig:triad_lam} in combination with the results for $\lambda=6$ and $\lambda=5$ (from experiment 1A) for a better comparison. Note that the extents of the FOV are different for experiment 1B. Therefore, in order to be consistent across different FOV sizes, the triadic energy exchanges are integrated between $0.5 < x/D <3D$ and $0<y/D<0.75D$. Note that as we increase $\lambda$ from 5 to 6, there is a clear transition in the energy exchange pathway. For $\lambda=5.5$, the pattern looks exactly similar to that for $\lambda=6$. For $\lambda=5.3$, the pattern appears to be at an intermediate state between that for $\lambda=5$ and $\lambda=5.5$. From $\lambda=5.5$, the energy exchange pattern remains qualitatively similar as we increase $\lambda$ and the magnitude of the energy transfers increases showing a stronger non-linear interaction between the modes at higher $\lambda$.

\begin{figure}
  \centerline{
  \includegraphics[clip = true, trim = 0 0 0 0 ,width= 1\textwidth]{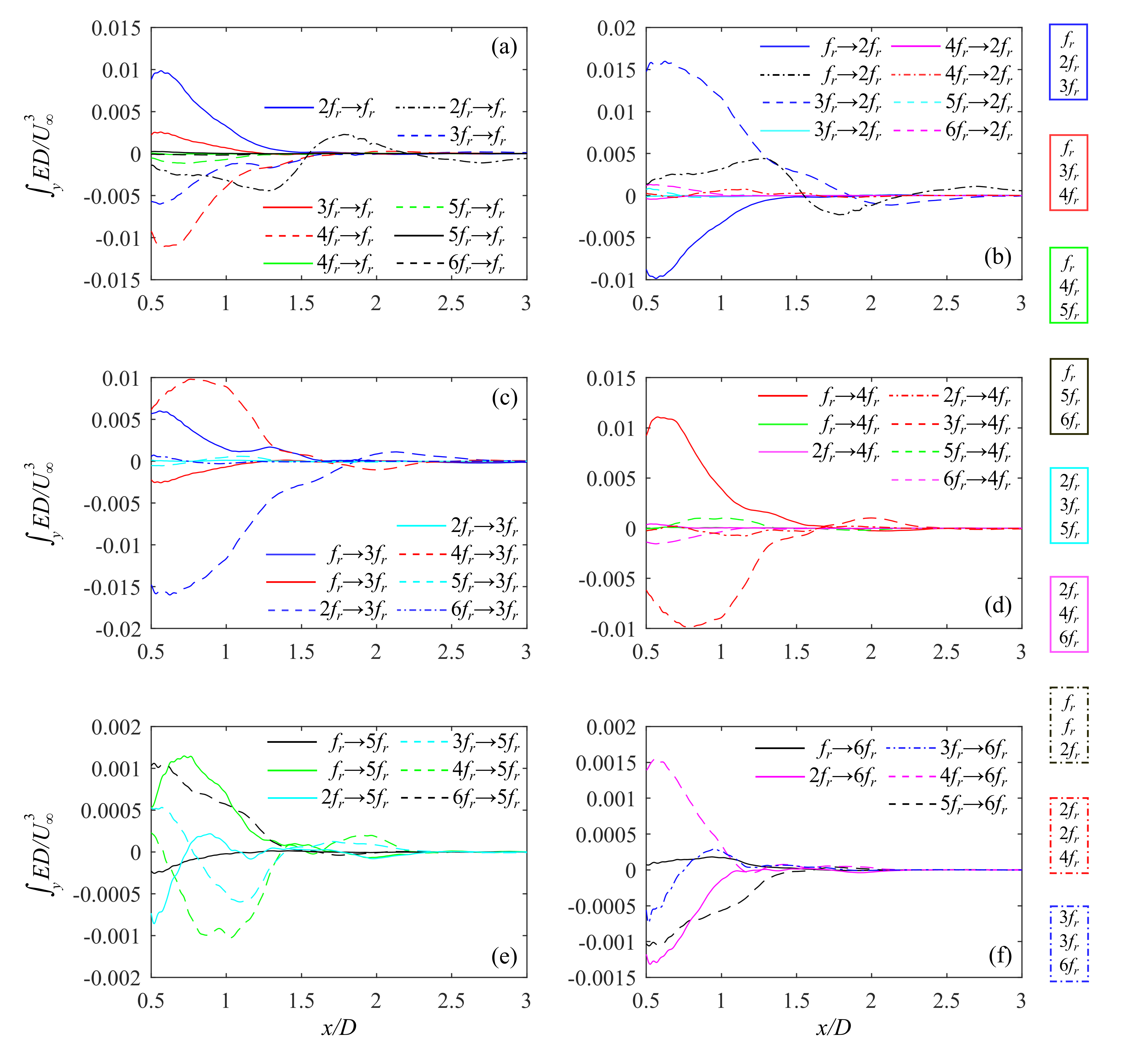}  }
 \caption{Energy transfers from different frequencies in different triads to (a) $f_r$, (b) $2f_r$, (c) $3f_r$, (d) $4f_r$, (e) $5f_r$ and (f) $6f_r$. The possible traids are shown on the right.}
\label{fig:triad}
\end{figure}

\rvvv{From figure \ref{fig:triad_full}(a), we can see that for $\lambda=6$, $2f_r$ gets most of its energy from $3f_r$.} However, $2f_r$ and $3f_r$ can form 2 triads together, one with $f_r$, and another with $5f_r$, hence it is not clearly known which triad is more important. We thus further split the inter-frequency energy transfers shown in figure \ref{fig:triad_full} into contributions from different triads. For brevity, we present the results only for $\lambda=6$ as it showed stronger non-linear interactions. In figure \ref{fig:triad} we show the streamwise evolution of the spanwise averaged inter-frequency energy transfers that correspond to different triads. Note that there are a total of 9 triads involving the frequencies $f_r - 6f_r$. These are shown in boxes of different colours and line types in figure \ref{fig:triad} for an easier interpretation. The first six triads involve 3 different frequencies and are shown in boxes with a solid line, let's call them triad type $I$. The last three triads have a repeated frequency and are shown in boxes with a dash-dotted line to distinguish them. These triads are termed triad type $II$. We first show the energy transfers to $f_r$ from the other frequencies in different triads in figure \ref{fig:triad}(a). Note that $f_r$ forms 4 triads of type $I$ and 1 triad of type $II$. For each triad of type $I$, it is possible to have two energy transfers to $f_r$ from the other two frequencies involved in the triad. For triads of type $II$, there will be only one transfer. Accordingly, we have a total of 9 transfers to $f_r$ as indicated in figure \ref{fig:triad}(a). The transfers corresponding to the first triad (shown in the blue solid box) are shown with blue lines. The transfer from the lower of the two frequencies of the triad (\textit{i.e.} $2f_r$) is shown in a solid blue line and that from the higher frequency ($3f_r$) is shown in a dashed blue line. The transfer from the $7th$ triad (shown in black dash dotted box) is shown in a black dash-dotted line. This same convention is used throughout to avoid confusion. Note that for $f_r$, only three triads 1, 2 and 7 are important. As a whole, $f_r$ loses some energy to $2f_r$, $3f_r$ and $4f_r$ through different triads. The rest of the transfers are small in comparison. For $2f_r$, only triad 1 and 7 are found to be important and both positively energise $2f_r$. \rvvv{The main source of energy for $2f_r$ is however $3f_r$ from triad 1 as discussed earlier}. For $3f_r$, only triad 1 and 2 are important. As a whole, $3f_r$ loses energy, most of which goes to energise $2f_r$. For $4f_r$, only the second triad is important. The energy transfers involving $5f_r$ and $6f_r$ are an order of magnitude smaller and can be ignored.

\begin{figure}
  \centerline{
  \includegraphics[clip = true, trim = 0 0 0 -20 ,width= 0.85\textwidth]{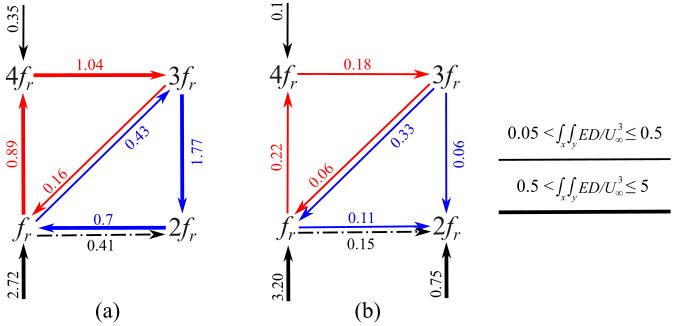}  }
 \caption{Triadic energy transfer pathways among the modes $f_r - 4f_r$ for (a) $\lambda=6$ and (b) $\lambda=5$. The line types and the colour of the arrows showing inter frequency energy transfers are consistent with that used to represent triads in figure \ref{fig:triad}. The solid black arrow shows the positive contribution from the $\tilde{P}_l$ term in equation \ref{eqn:coherent_TKE}. The thickness of the arrows vary according to the magnitude of the energy transfers as indicated.  }
\label{fig:triad_sch}
\end{figure}

\rvvv{The above analysis allows us to drastically simplify the network of non-linear energy exchanges by using a fewer number of triads. To be specific, for $\lambda=6$, we can only retain the triads 1, 2 and 7 and discard the rest. This approximated network of energy transfer is schematically shown in figure \ref{fig:triad_sch}(a) for $\lambda=6$ and it summarises the key energy exchanges in the tip vortex system discussed above. The energy exchanges in triad 1 and 2 are shown in blue and red solid lines respectively, while the black dash dotted line represents the energy transfer in triad 7, similar to the convention used in figure \ref{fig:triad}. Note that triads 1 and 2 show a cyclic nature of energy transfers and form a net inverse energy cascade (\textit{i.e.} energy transfer from high to low frequency), while triad 7 shows a forward cascade. For a comparison, the positive contributions from the mean flow production term ($\tilde{P}_l$) in equation \ref{eqn:coherent_TKE} are also shown by solid black lines. The numbers show the magnitudes of the energy transfers and the same is also highlighted by the thicknesses of the arrows. In both triads 1 and 2, the direct energy transfers between $f_r$ and $3f_r$ are weaker. In triad 1, energy primarily flows from $3f_r$ towards $f_r$ via the secondary mode $2f_r$ which is a nice representation of the tip vortex merging process \citep{felli2011mechanisms}. In triad 2, the dominant energy transfer direction is the opposite, \textit{i.e.} from $f_r$ to $3f_r$ via $4f_r$. }

\rvvv{A similar schematic of energy transfers among the modes $f_r - 4f_r$ for $\lambda=5$ is shown in figure \ref{fig:triad_sch}(b). Note that for $\lambda=5$ a simplification of the energy transfer network is not readily possible as all the modes exhibit energy transfers of similar magnitudes to/from them (see figure \ref{fig:triad_full}(b)). Therefore, the purpose of figure \ref{fig:triad_sch}(b) is solely to compare the triads 1, 2 and 7 between $\lambda=5$ and $\lambda=6$. First of all, the energy transfers are much weaker for $\lambda=5$. Secondly, although triad 2 is still cyclic in nature, triad 1 has become non-cyclic. The primary change has occurred around $2f_r$ which is no longer strongly energised by $3f_r$. Triads 1 and 7 contribute to some non-linear energy gain of $2f_r$ but it gets most of its energy from the mean flow.}

\rvvv{These energy exchanges can elucidate the merging process of the tip vortices. Recently, \citet{biswas2024effect} reported two different merging processes for two different $\lambda$s using the same rotor model. For the lower $\lambda$ ($\lambda=4.5$) the merging of the tip vortices resembled a two-step process where first $2f_r$ was formed and $f_r$ became energetic further downstream, as also reported by \citet{felli2011mechanisms, sherry2013interaction}. Contrastingly, for the higher $\lambda$ ($\lambda=6$), three tip vortices appeared to combine almost directly in what they referred to as a one-step process (see also supplementary video 2 of \citet{biswas2024effect}). For the same $\lambda$, we however do not see a direct energy transfer from $3f_r$ to $f_r$ in figure \ref{fig:triad_sch}(a), as one might expect from a one-step process. The energy transfer still happens in two steps, first from $3f_r$ to $2f_r$ and then from $2f_r$ to $f_r$ as predicted by \citet{felli2011mechanisms}. However, due to the fact that $f_r$ and $2f_r$ attain their maximum energy at almost the same streamwise location (see figures \ref{fig:mode_tip}(a) and (b)), we can say that these two steps take place almost concurrently, making it look like a one-step process. For $\lambda=5$, there is a larger streamwise separation between the points where $f_r$ and $2f_r$ attain their maximum energy, visually indicating a two-step merging process. The increased separation between the helices at the lower $\lambda$, however, results in a much weaker non-linear interaction. Furthermore, the injection of energy from the mean flow to $2f_r$ results in a reorganisation of the energy transfer pattern in triad 1, \textit{i.e.} from a cyclic to a non-cyclic one.  }

\section{The near wake and wake recovery}

For a wind turbine it is important to quantify the extent of the near wake, particularly in the context of designing wind farm layouts as the near wake contains energetic coherent structures capable of inducing fatigue damage to the subsequent turbines. The extent of the near 
    wake depends on several factors such as the nature and intensity of the freestream turbulence level, turbine geometry, operating condition and so on and hence it is often vaguely defined between 2-4 rotor diameters downstream of the rotor \citep{vermeer2003wind, foti2016wake}. Several attempts have been made to quantify the near wake for instance by observing self-similarity in the time-averaged wake profile \citep{sorensen2015simulation} or by looking at the variation of time-averaged coherent or turbulent kinetic energy in the wake \citep{de2021pod, gambuzza2023influence}. In contrast, \citet{biswas2024effect} endorsed a more dynamic point of view and defined the extent of the near wake as the location where the strength of wake meandering or $f_{wm}$ (dominant dynamic feature in the far wake) surpassed that of the tip vortices or $3f_r$ (dominant frequency in the near wake). They also defined a convective length scale, $L_c = \pi D/\lambda$ \rv{(which can be physically interpreted as the distance travelled at the freestream velocity in the time taken for one complete rotation of the turbine)} and reported that the near wake location was $\approx 3L_c$ for a range of $\lambda$s tested, or in other words, the near wake's extent scaled with $1/\lambda$. \rvv{The strength of a frequency at a particular streamwise ($x$) location was however defined as the magnitude of the spectral peak at that particular frequency}. As the strengths of the frequencies depended only on $x$, the spatial distribution of the strength or energy associated with frequencies was not taken into account in the definition of the near wake. In this section we explore other ways to quantify the extents of the near/far wake based on the knowledge gained earlier about the nature of the different modes and the energy exchanges to/from them. 

\begin{figure}
  \centerline{
  \includegraphics[clip = true, trim = 0 0 0 0 ,width= 0.85\textwidth]{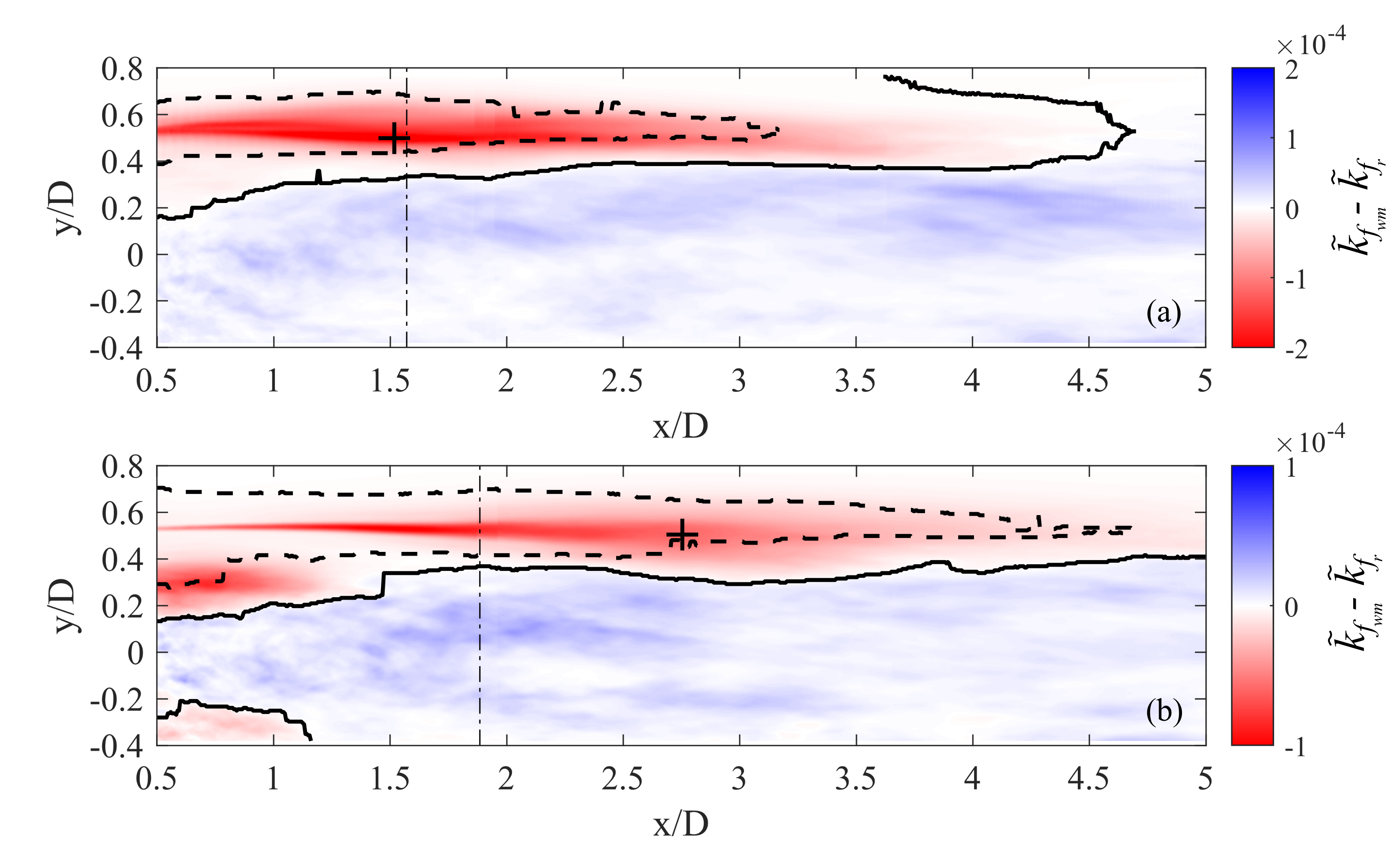}  }
 \caption{Filled contours showing relative kinetic energy of the wake meandering mode with respect to the mode associated with $f_r$ (denoted by $\tilde{k}_{f_{wm}} - \tilde{k}_{f_r}$) for (a) $\lambda=6$ and (b) $\lambda=5$. The solid and dashed black lines correspond to $\tilde{k}_{f_{wm}} - \tilde{k}_{f_r}=0$ and $\tilde{k}_{f_{wm}} - \tilde{k}_{3f_r}=0$ respectively. The \textbf{+} sign shows the location where the kinetic energy corresponding to $f_r$ is maximum. The dash-dotted vertical line shows the streamwise location corresponding to 3 convective length scales or $3L_c$ defined in \citep{biswas2024effect}. }
\label{fig:wake_boundary}
\end{figure}

 We first obtain the time-averaged kinetic energy of the modes associated with $f_{wm}$ ($k_{f_{wm}}$) and $f_r$ ($k_{f_r}$) and plot the relative kinetic energy $k_{f_{wm}} - k_{f_r}$ for $\lambda = 6$ and $\lambda=5$ in figures \ref{fig:wake_boundary}(a) and \ref{fig:wake_boundary}(b) respectively. The solid black line corresponding to $k_{f_{wm}} - k_{f_r} = 0$ nicely distinguishes the regions where $f_{wm}$ or $f_r$ is stronger than the other. Similarly, the dashed black line shows the contour corresponding to $k_{f_{wm}} - k_{3f_r} = 0$ which is always found to be located within the contour traced by $k_{f_{wm}} - k_{f_r} = 0$ for the $\lambda$s tested. Similar contours can be obtained using the harmonics of $f_r$ and $3f_r$, but they are generally weaker in nature. We therefore propose that $k_{f_{wm}} - k_{f_r} = 0$ can be used to distinguish the inner wake involving low-frequency dynamics from the outer wake that contains high-frequency modes related to the tip vortices. The earlier and stronger tip vortex merging process for $\lambda=6$ results in a quicker disintegration of the tip vortex system. The outer wake thus vanishes at $x/D \approx 4.7$ creating a path for rapid exchange of mass and momentum between the inner wake and the non-turbulent background fluid in this case, aiding wake recovery. We therefore argue that the end of the outer wake can be considered as the initiation of the far wake where wake meandering is the only discernible frequency signature. The extent of the outer wake and hence the far wake depends on $\lambda$. For $\lambda=5$, the outer wake extends beyond $x/D =5$, therefore we can expect the far wake to also scale with $1/\lambda$ at least in the presence of low inflow turbulence, similar to the near wake \citep{biswas2024effect}. 

\begin{figure}
  \centerline{
  \includegraphics[clip = true, trim = 0 0 0 0 ,width= \textwidth]{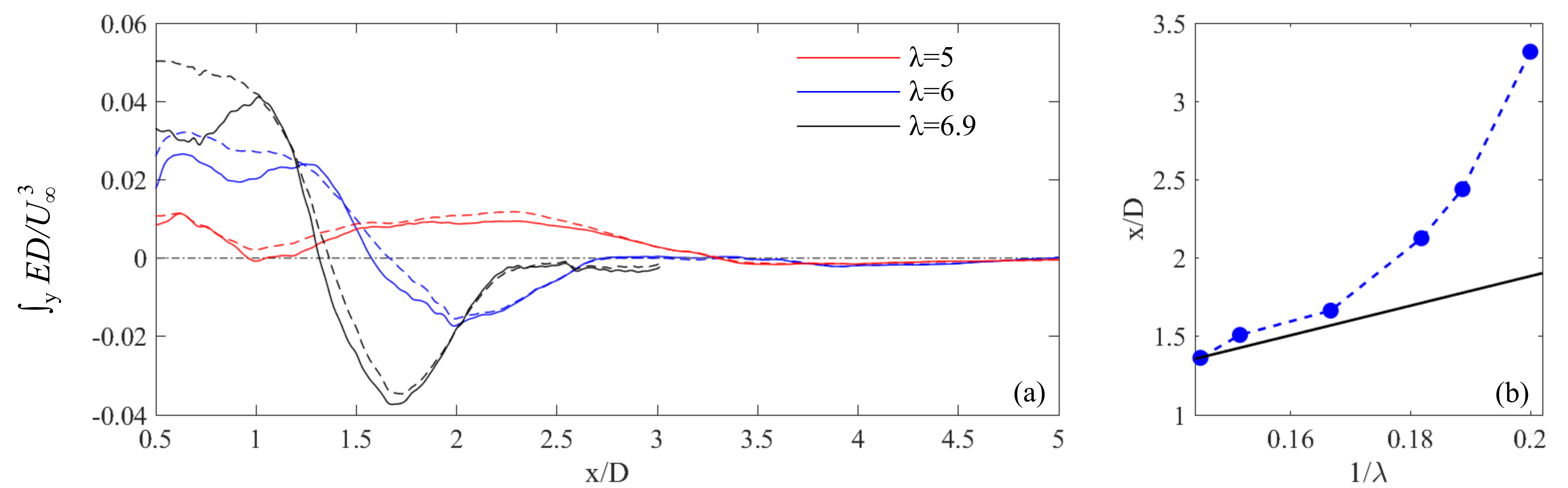}  }
 \caption{(a) Streamwise variation of spanwise averaged mean flow production term for the tip vortex system (shown by the solid lines) and $f_r$ (shown by the dashed line) for diffrerent $\lambda$s. (b) The variation of the location where wake recovery initiates or $x_{wr}$ with $\lambda$ (\textbf{\textcolor{blue}{--$\sbullet[1.2]$--}}). The solid black line shows the streamwise location $x/D = 3L_c$. }
\label{fig:near_wake}
\end{figure}

Next we look at the evolution of the mean flow production term ($\tilde{P}_l$) of the tip vortex system. In figure \ref{fig:near_wake}(a) we show the streamwise evolution of the spanwise averaged (along $y$) mean flow production term for the tip vortex system ($\sum_{l=f_r}^{l=6f_r} \tilde{P}_l$) for 3 different $\lambda$s (by the solid lines). While the dashed lines show the mean flow production term only for $f_r$ ($\tilde{P}_{f_r}$) for the corresponding $\lambda$s. Note that $\tilde{P}_{f_r}$ is close to $\sum_{l=f_r}^{l=6f_r} \tilde{P}_l$ for all $\lambda$s which implies that most of the energy exchanges between the mean flow and the tip vortex system happens through $f_r$. As discussed earlier, disintegration of the tip vortex system is essential to re-energise the wake. Note that close to the rotor $\sum_{l=f_r}^{l=6f_r} \tilde{P}_l > 0$, \textit{i.e.} the tip vortex system first draws energy from the mean flow as a whole. However, after some distance downstream it starts to transfer the energy back to the mean flow as indicated by $\sum_{l=f_r}^{l=6f_r} \tilde{P}_l <0$. The streamwise location where the combined mean flow production term for the tip vortex system changes sign can be ascribed as the point of initiation of wake recovery (let us denote it by $x_{wr}$). Note that as $\lambda$ increases $x_{wr}$ moves closer to the rotor. Moreover, an estimation of $x_{wr}$ can be obtained just by looking at the production term of $f_r$ which again highlights the importance of $f_r$ in the distinction of near from far wake. In figure \ref{fig:near_wake}(b) we show the variation of $x_{wr}$ obtained from $\tilde{P}_{f_r}$ with $\lambda$. The extent of the near wake, $\approx 3L_c$, proposed by \citet{biswas2024effect} is also shown for a comparison. $x_{wr}$ shows a nice trend with $\lambda$. More interestingly, for higher $\lambda$, $x_{wr}$ becomes close to $3L_c$, further highlighting the efficacy of $L_c$ as a length scale in the near wake.

%\citet{foti2016wake} constructed 3 dimensional spatio-temporal wake meander profiles. Between near and far wake region ($2D-4D$), they found an increase in the meander amplitude and the meander profiles approached the tip vortices radially. 

\begin{figure}
  \centerline{
  \includegraphics[clip = true, trim = 0 0 0 0 ,width= 0.65\textwidth]{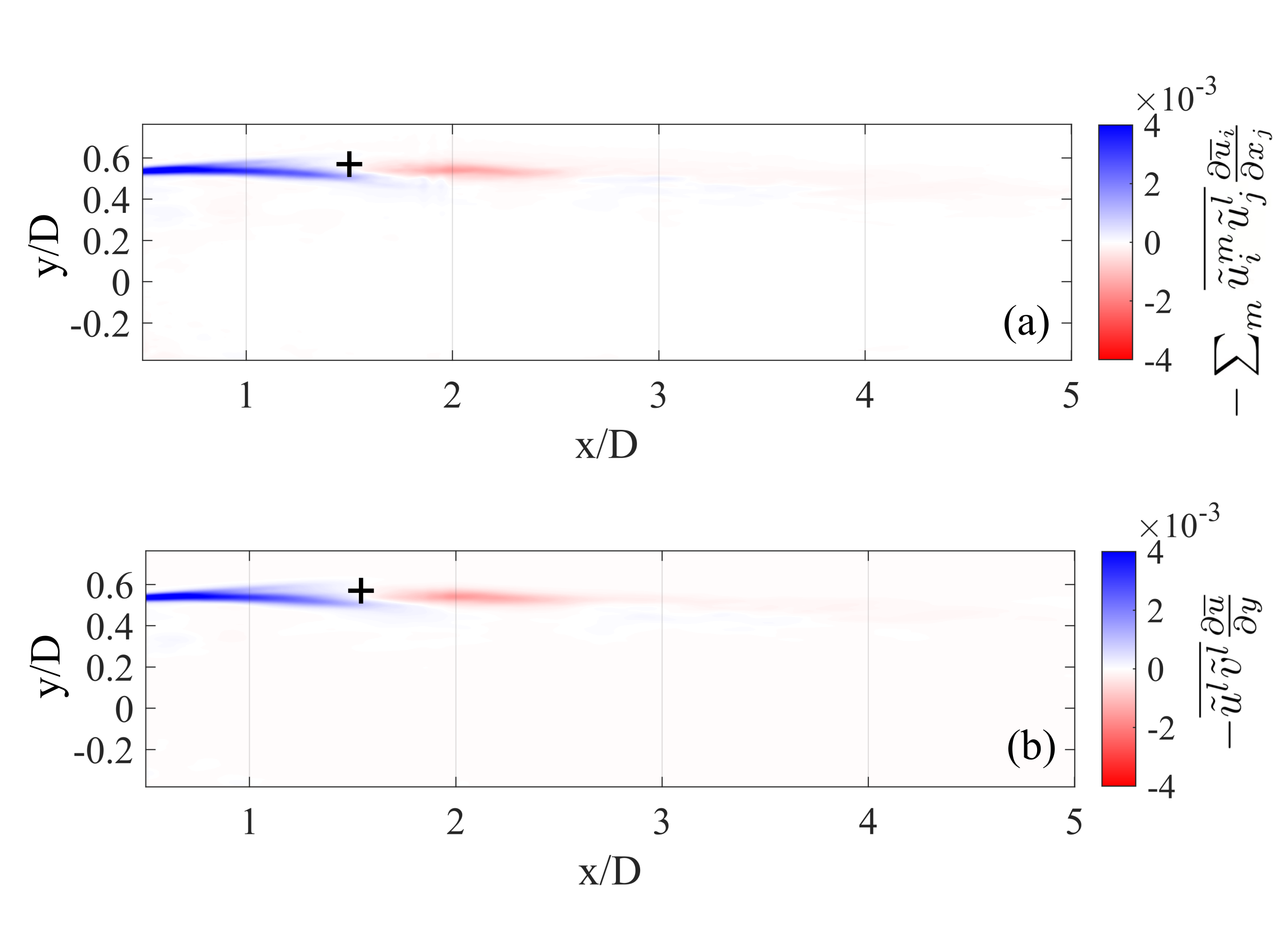}  }
 \caption{(a) Mean flow production term ($\tilde{P}_l$) of $f_r$ for $\lambda=6$, (b) shows the leading term of $\tilde{P}_l$. The `\textbf{+}' sign marks the location of the sign change. }
\label{fig:Pl_approx}
\end{figure}

\rv{To understand why the $\tilde{P_l}$ term for $f_r$ changes sign after some distance downstream, let us take a closer look at its composition. \citet{baj2017interscale} defined the term as $\tilde{P}_l = -\sum_m \overline{\tilde{u}^m_i \tilde{u}^l_j} \frac{\partial \overline{u}_i}{\partial x_j}$, where $m$ can be any coherent mode selected in the reduced order representation of the flow including the $l_{th}$ mode. \rvv{Assuming that the velocity components of the different modes are uncorrelated, \textit{i.e.} $\overline{ \tilde{u}^l_i \tilde{u}^m_j} \approx 0$ for any $l \neq m$, we can say that $\tilde{P}_l \approx -\overline{\tilde{u}^l_i \tilde{u}^l_j} \frac{\partial \overline{u}_i}{\partial x_j}$.} Next, we can show that the transverse gradient of the streamwise velocity ($\frac{\partial \overline{u}}{\partial y}$) is an order of magnitude stronger than the other gradients, at least in the tip shear layer region so we can further approximate the term as $\tilde{P}_l \approx -\overline{\tilde{u}^l \tilde{v}^l} \frac{\partial \overline{u}}{\partial y}$. In figure \ref{fig:Pl_approx}(a) we show the $\tilde{P}_l$ (without any approximation) field for $f_r$. As can be expected, $\tilde{P}_l$ is concentrated only in the tip shear layer region and the `\textbf{+}' sign shows the location where the term changes sign, \textit{i.e.} the location of the onset of wake recovery ($x_{wr}$). Figure \ref{fig:Pl_approx}(b) shows the $\tilde{P}_l$ field only with the dominant term ($-\overline{\tilde{u}^l \tilde{v}^l} \frac{\partial \overline{u}}{\partial y}$) and it looks almost exactly similar to figure \ref{fig:Pl_approx}(a), hence justifying the approximation.}

\rv{Now, $\frac{\partial \overline{u}}{\partial y}$ is always positive in the tip shear layer region. The observed change of sign in $\tilde{P}_l$ thus requires a change in the nature of correlation (negative to positive) between the streamwise and transverse velocity components of $f_r$. This is similar to the observation of \citet{lignarolo2015tip} for a two-bladed turbine. They argued that the net transport of kinetic energy towards the wake centreline (the component contributing to the wake recovery) due to the periodic motions can be given by the gradients of the kinetic energy flux $-\overline{\tilde{u}^l \tilde{v}^l}\overline{u}$. They further showed that at a certain distance downstream, the orientation/inclination of the tip vortex pair undergoing merging with respect to the streamwise direction changed from negative ($<90^o$) to positive ($>90^o$), resulting in a change in the correlation between the streamwise and transverse velocity component (see figure 17 of \citet{lignarolo2015tip}). From figure \ref{fig:sch} here and also from supplementary video 2 of \citet{biswas2024effect} we can see that for $\lambda=6$, the vortex triplet, undergoing merger, changes its inclination at $x/D \approx 1.5$. Consequently, we observe a sign change in the $\tilde{P}_l$ term for $f_r$ at $x/D \approx 1.5$ for $\lambda=6$.}

\section{Conclusion}
The near wake of a wind turbine is abundant with coherent structures. Studying the interaction between these coherent structures is a key to understanding the spatio-temporal evolution of the wake and phenomena like wake recovery. We conducted particle image velocimetry (PIV) experiments to identify and extract the coherent structures in the near wake of a rotor model having a nacelle and a tower, thereby making it representative of a utility-scale wind turbine for two main tip speed ratios ($\lambda$), $\lambda=6$ and $\lambda=5$. The coherent structures were identified using a multiscale triple decomposition of the velocity field using optimal mode decomposition or OMD \citep{wynn2013optimal, baj2015triple}. A large number of high-frequency modes were extracted related to the tip vortices including modes with a frequency equal to the turbine's rotational frequency ($f_r$), blade passing frequency ($3f_r$) and their harmonics ($2f_r$ and $4f_r - 6f_r$ ) as well as a number of low-frequency modes which included sheddings from the tower ($f_T$), the nacelle ($f_n$) and wake meandering ($f_{wm}$). The spatial nature and strength of the modes depended on $\lambda$. For the higher $\lambda$ ($\lambda=6$), the modes related to the tip vortices were energetic closer to the rotor, owing to the early interaction between the tip vortices for a higher $\lambda$ \citep{sherry2013interaction}. For the low-frequency modes, the modes associated with $f_n$ and $f_T$ were similar for the two $\lambda$s. Interestingly however, the wake meandering mode was found to be more energetic for the higher $\lambda$ consistent with the observations of \citet{biswas2024effect}. 

Further insights are gained about the nature of the different modes by studying the energy exchanges to and from them by using the multiscale triple decomposed coherent kinetic energy (CKE) budget equation derived by \citet{baj2017interscale}. The low-frequency modes associated with the sheddings from the tower and the nacelle as well as the wake meandering mode were found to be primarily energised by the mean flow, hence acting similarly to a `primary mode', as termed in previous studies \citep{baj2017interscale, biswas2022energy}. \rvv{The tip vortex system} on the other hand was found to have a variety of energy sources such as energy production from the mean flow, non-linear triadic energy production or both. \rvv{The mode associated with $f_r$ behaved like a primary mode for both $\lambda$s, while the nature of $2f_r$ changed with $\lambda$. For $\lambda=6$, $2f_r$ gained most of its energy through the non-linear triadic energy production term in the CKE budget equation, hence it is akin to a `secondary mode' \citep{baj2017interscale, biswas2022energy}. For $\lambda=5$ on the other hand, the mode received most of its energy through the mean flow production term, while having a non-negligible positive contribution from the triadic interaction term, therefore acting like a `mixed mode' as discussed in \citet{biswas2022energy}. The differences in the energy exchanges observed for the higher and lower $\lambda$ were shown to be consistent with the `one-step' and `two-step' merging processes reported earlier by \citet{biswas2024effect}.} Some triadic interaction was observed between the low frequency modes as well, but its net contribution was small compared to the energy production from the mean flow.

\begin{figure}
  \centerline{
  \includegraphics[clip = true, trim = 0 0 0 0 ,width= \textwidth]{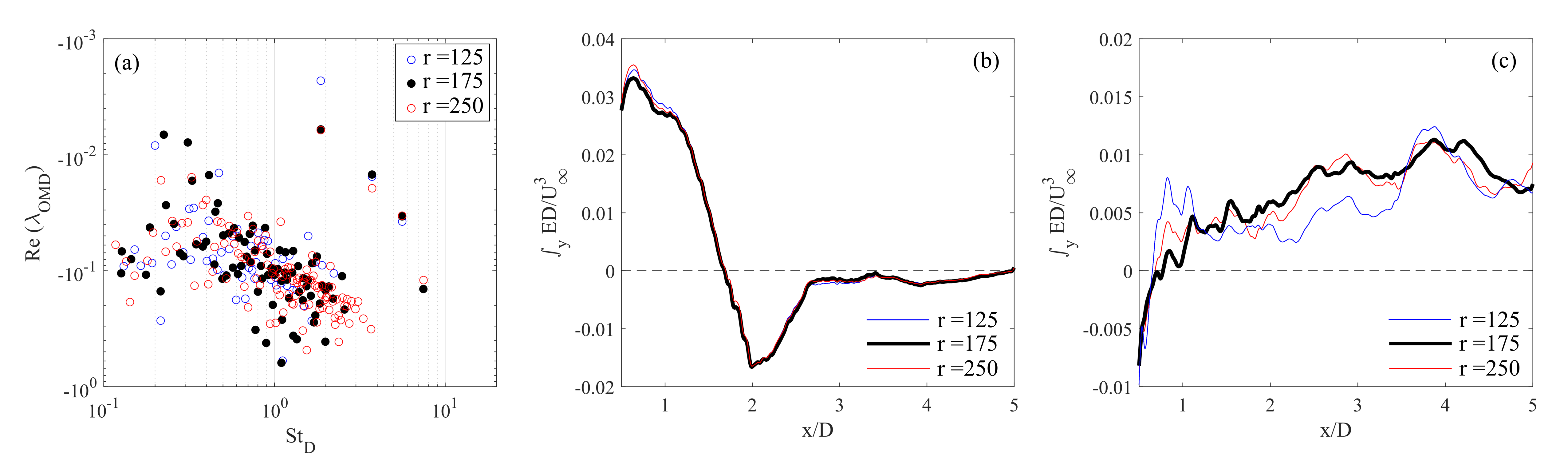}  
 
  }
 \caption{(a) OMD spectra for different ranks $r$. (b) and (c) show the streamwise evolution of the spanwise-averaged $\tilde{P}_l$ term for $f_r$ and $f_{wm}$ respectively for different $r$. }
\label{fig:conv}
\end{figure}

A complex network of triadic energy exchanges between the modes associated with $f_r - 6f_r$ is identified and discussed. For $\lambda=6$, energy was found to flow from $3f_r$ to $2f_r$ and then from $2f_r$ to $f_r$ via \rv{the non-linear triadic interaction term} in the triad formed by $f_r, 2f_r$ and $3f_r$, similar to the observation of \citet{felli2011mechanisms}. In the triad formed by $f_r,3f_r$ and $4f_r$, $f_r$ transfered some energy to $4f_r$. $4f_r$ then transfered almost the same amount of energy to $3f_r$, resulting in a much weaker but still dynamically important $4f_r$ mode. The same pattern of triadic energy exchanges was observed for 4 different $\lambda$s ($\lambda=5.5$, $\lambda=6$, $\lambda=6.6$ and $\lambda=6.9$). The triadic energy exchanges for the lower $\lambda$s were found to be much weaker due to the increased separation between the tip vortex filaments. 

Finally, attempts are made to identify the boundaries between the inner/outer wake and near/far wake based on the modes and their energy contents. It is shown that the inner wake (involving low-frequency dynamics) can be distinguished from the outer wake (involving high-frequency structures) by comparing the kinetic energy associated with the turbine's rotational frequency $f_r$ and wake meandering ($f_{wm}$). To identify the location of the onset of wake-recovery (defined as $x_{wr}$) we looked at the combined mean flow production term of the modes related to the tip vortices ($f_r - 6f_r$). Initially, the tip vortex system is found to extract energy from the mean flow, but further downstream, the combined production term became negative, implying a net energy transfer back to the mean flow. More interestingly, most of the energy exchange between the mean flow and the tip vortex system was found to happen through the turbine's rotational frequency. \rv{The sign change in the mean flow production term of $f_r$ was shown to be related to a switch in the nature of correlation between the streamwise and transverse velocity components of the $f_r$ mode resulting from the change of inclination of the tip vortex system undergoing merging, similar to the observation of \citet{lignarolo2015tip} for a two-bladed rotor}.

\begin{figure}
  \centerline{
  \includegraphics[clip = true, trim = 0 0 0 0 ,width= \textwidth]{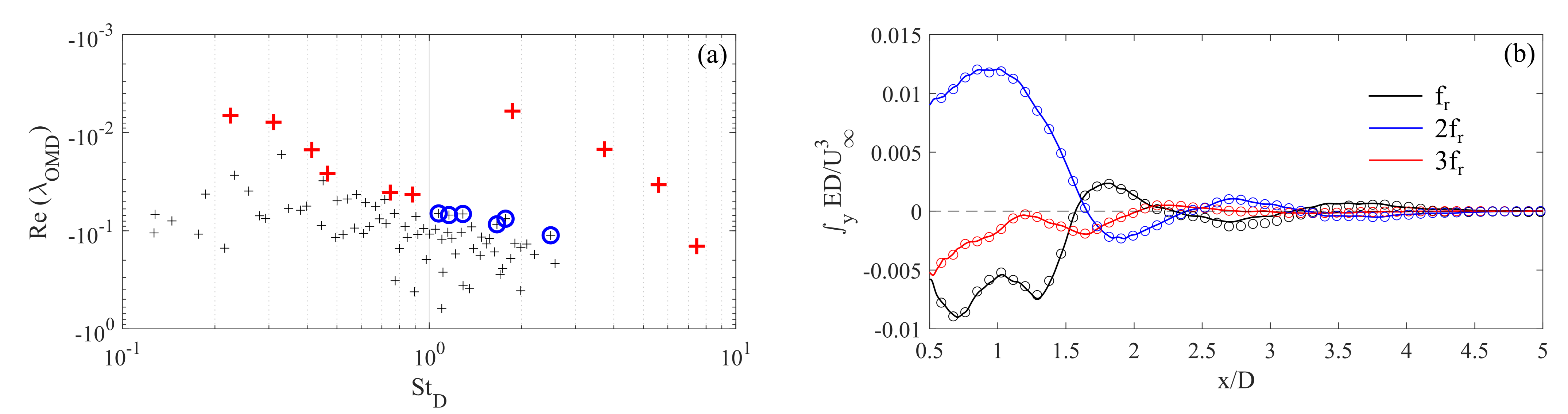}  
 
  }
 \caption{\rvv{(a) OMD spectrum for $\lambda=6$ from experiment 1A. The `\textbf{\textcolor{blue}{$\circ$}}' symbols show the additional modes retained in the range $1\lesssim St_D \lesssim3$. (b) shows the streamwise evolution of the spanwise-averaged net triadic energy production term for $f_r - 3f_r$ with (circles) and without (solid lines) the inclusion of the additional modes.} }
\label{fig:hi_f}
\end{figure}

\section*{Appendix 1: Convergence of the OMD modes}
To estimate the possible effects of the rank $r$ of the OMD matrices (which needs to be selected \textit{a priori}), the decomposition is performed with different $r$s. OMD is first applied to the data set from experiment 1A for $\lambda=6$ for different $r$s. The resultant spectra are shown in figure \ref{fig:conv}(a). For $r=125$, only the modes from $f_r - 3f_r$ are captured. While, for the higher ranks, $4f_r$ is also captured. The dominant wake meandering frequency around $St_D\approx 0.2$ was captured for all the $r$s as well as frequencies in the range \rvv{$0.4\lesssim St_D \lesssim 0.5$ and $0.7\lesssim St_D \lesssim 0.9$} which are associated with the vortex sheddings from the nacelle and the tower respectively. To evaluate the effect of $r$ on the energy budget analysis, we look at the primary source term for $f_r$ and $f_{wm}$, \textit{i.e.} the coherent energy production term ($\tilde{P}_l$) for different $r$s. The streamwise evolution of the spanwise averaged $\tilde{P}_l$ term for the frequencies are shown in figure \ref{fig:conv}(b) and \ref{fig:conv}(c) for different $r$s. For $f_r$, there was no significant variation in $\tilde{P}_l$ with changes in $r$. A similar observation was noted for all the budget terms for all the tip vortex related modes. The $\tilde{P}_l$ term for $f_{wm}$ is found to be more sensitive to the selection of $r$, especially for $r=125$, the term looks significantly off from the other two curves in figure \ref{fig:conv}(c). This can be expected as the energy budget terms were much noisier for the low-frequency modes. Nevertheless, the energy budget terms do not change significantly when summed over the entire domain, especially after $r=175$. \rv{Accordingly, we only present the results for $r=175$}. \\

\section*{Appendix 2: Selection of OMD modes}

\rvv{The OMD spectrum in figure \ref{fig:spectra} shows a number of modes with characteristic frequencies in the range $1\lesssim St_D \lesssim3$. The modes could be related to non-linear interactions between the $f_r$ mode and one of the low-frequency modes (the wake meandering mode or the sheddings from the nacelle or the tower). These modes were, however, not selected in the reduced order representation of the flow due to their highly-damped nature. Nevertheless, we repeated the analysis by including some of these modes for $\lambda=6$ to see if there were any changes in the energy exchange patterns observed, especially in the non-linear triadic energy exchanges. The OMD spectrum for $\lambda=6$ (with $r=175$) is reproduced in figure \ref{fig:hi_f}(a). The originally selected modes are shown by a `\textbf{\textcolor{red}{+}}' and the extra modes selected in the range $1\lesssim St_D \lesssim 3$ are marked by a `\textbf{\textcolor{blue}{$\circ$}}' sign. In figure \ref{fig:hi_f}(b) we show the streamwise evolution of the spanwise averaged net triadic energy production (the $\tilde{T}_l^+ - \tilde{T}_l^- $ term in equation \ref{eqn:coherent_TKE}) for $f_r$, $2f_r$ and $3f_r$ obtained by using fewer modes (shown by solid lines) and by including the additional modes in the range $1\lesssim St_D \lesssim3$ (using circles). Inclusion of the additional modes clearly does not alter the triadic energy exchanges in the tip vortex system, indicating negligible non-linear interaction between the tip vortex system and the low-frequency modes, at least within the field of view considered.   } \\

%\begin{figure}
%  \centerline{
%  \includegraphics[clip = true, trim = 0 0 0 0 ,width= \textwidth]{figures/domain_dependence.png}  
 
%  }
% \caption{(a) OMD spectra for different ranks $r$. (b) and (c) show the streamwise evolution of spanwise averaged $\tilde{P}_l$ term for $f_r$ and $f_{wm}$ respectively for different $r$. }
%\label{fig:domain}
%\end{figure}

%\textcolor{black}{\rule{1cm}{1pt}} $\tilde{P}_l$ \textcolor{red}{\rule{1cm}{1pt}} $-\tilde{C}_l$ \textcolor{blue}{\rule{1cm}{1pt}} $\tilde{T}^+_l - \tilde{T}^-_l$ \textcolor{green}{\rule{1cm}{1pt}} $-\hat{P}_l$ \textcolor{orange}{\rule{1cm}{1pt}} $\tilde{D}_l$ \textcolor{magenta}{\rule{1cm}{1pt}} $-\tilde{\epsilon}_l$ \textcolor{black}{\rule{1cm}{1pt}} $\tilde{\zeta}_l$

\noindent \textbf{Funding}. NB gratefully acknowledges funding through the Imperial College London President's Scholarship and the Engineering and Physical Sciences Research Council (EPSRC) through grant EP/T51780X/1. OB gratefully acknowledges funding from EPSRC through grant no. EP/V006436/1.\\

 \noindent \textbf{Declaration of interests}. The authors report no conflict of interest.\\

 \noindent For the purposes of open access, the authors have applied a Creative Commons Attribution (CC BY) licence to any Author Accepted Manuscript (AAM) version arising.

\bibliography{bib}

\end{document}